\documentclass[letterpaper,twocolumn,10pt]{article}
\usepackage[dvipsnames]{xcolor}
\usepackage{usenix-2020-09}

\usepackage{xurl} 

\usepackage{tikz}
\usepackage{amsmath}

\usepackage[dvipsnames]{xcolor}
\usepackage{mdframed}

\mdfsetup{
  linecolor=Maroon,          
  linewidth=1pt,             
  frametitlerule=true,       
  frametitlebackgroundcolor=gray!20, 
  frametitlerulewidth=0.5pt, 
  frametitlefont=\bfseries,  
  frametitlealignment=\raggedright, 
}

\usepackage{enumitem}
\setlist[enumerate]{noitemsep,topsep=2pt,leftmargin=*}
\setlist[itemize]{noitemsep,topsep=2pt,leftmargin=*}

\usepackage{fontawesome}

\newcommand{\myCheck}{\textcolor{OliveGreen}{\faCheck}}
\newcommand{\myCheckOrange}{\textcolor{orange}{\faCheck}}
\newcommand{\myCross}{\textcolor{red}{\faTimes}}
\newcommand{\myStar}{{\color{red}$^*$}}

\newcommand{\tightpar}[1]{{\smallskip \noindent\bf #1}}

\usepackage{longtable}
\usepackage{caption}

\usepackage{multicol}
\usepackage{multirow}

\usepackage{color}
\usepackage{listings}

\definecolor{editorGray}{rgb}{0.95, 0.95, 0.95}
\definecolor{editorOcher}{rgb}{1, 0.5, 0} 
\definecolor{editorGreen}{rgb}{0, 0.5, 0} 
\definecolor{btq}{rgb}{0.03, 0.91, 0.87} 
\definecolor{dtq}{rgb}{0.0, 0.81, 0.82} 
\definecolor{cdb}{rgb}{0.37, 0.62, 0.63} 
\lstdefinelanguage{js}{
	keywords={typeof, new, true, false, try, catch, function, return, null, catch, switch, var, if, in, for, while, do, else, case, break,let, const, throw, with, await},
	keywordstyle=\color{Maroon}\bfseries,
	ndkeywords={class, export, boolean, throw, implements, this, async},
	ndkeywordstyle=\color{darkgray}\bfseries,
	identifierstyle=\color{black},
	sensitive=false,
	comment=[l]{//},
	morecomment=[s]{/*}{*/},
	commentstyle=\color{darkgray}\ttfamily,
	stringstyle=\color{OliveGreen}\ttfamily,
	escapeinside={/*\#}{\#*/},	
	morestring=[b]',
	morestring=[b]",
	morestring=[b]`
}
\newcommand{\tool}{\textsc{GHunter}}

\begin{document}

\date{} 

\title{\Large \bf \tool: Universal Prototype Pollution Gadgets in JavaScript Runtimes}

\author{
{\rm Eric Cornelissen}\\
KTH Royal Institute of Technology
\and
{\rm Mikhail Shcherbakov}\\
KTH Royal Institute of Technology
\and
{\rm Musard Balliu}\\
KTH Royal Institute of Technology
}

\maketitle

\begin{abstract}

Prototype pollution is a recent vulnerability that affects JavaScript code, leading to high impact attacks such as arbitrary code execution and privilege escalation.  The vulnerability is rooted in JavaScript's prototype-based inheritance, enabling attackers to inject arbitrary properties into an object's prototype at runtime. The impact of prototype pollution depends on the existence of otherwise benign pieces of code (gadgets), which inadvertently read from these attacker-controlled properties to execute security-sensitive operations. While prior works primarily study gadgets in third-party libraries and client-side applications, gadgets in JavaScript runtime environments are arguably more impactful as they affect any application that executes on these runtimes.

In this paper we design, implement, and evaluate a pipeline, \tool{}, to systematically detect gadgets in V8-based JavaScript runtimes with prime focus on Node.js and Deno. \tool{} supports a lightweight dynamic taint analysis to automatically identify gadget candidates which we validate manually to derive proof-of-concept exploits. We implement \tool{} by modifying the V8 engine and the targeted runtimes along with features for facilitating manual validation. Driven by the comprehensive test suites of Node.js and Deno, we use \tool{} in a systematic study of gadgets in these runtimes. We identified a total of 56 new gadgets in Node.js and 67 gadgets in Deno, pertaining to vulnerabilities such as arbitrary code execution (19), privilege escalation (31), path traversal (13), and more. Moreover, we systematize, for the first time, existing mitigations for prototype pollution and gadgets in terms of development guidelines. We  collect  a list of vulnerable applications  and revisit the fixes through the lens of our guidelines. Through this exercise, we also identified one high-severity  CVE leading to remote code execution, which was due to incorrectly fixing a gadget.

\end{abstract}

\section{Introduction}

JavaScript's widespread adoption as a go-to programming language for full-stack development speaks to its popularity, but it also exposes the applications to  
heightened security risks. Researchers and practitioners are well-aware of these issues, as witnessed by a multitude of prior studies ~\cite{StockJS017,ZimmermannSTP19,StaicuSBPS19,duantowards}. JavaScript runtime environments, such as Node.js ~\cite{Nodejs} and Deno ~\cite{Deno}, which lie at the heart of server-side JavaScript applications, become appealing targets for attackers \cite{BrownNWEJS17,StaicuPL18,AhmadpanahHBOS21,duantowards,Li21,Xiao21,ShcherbakovBS23}. Vulnerabilities in the runtime environments can compromise the security of applications running atop. In this paper, we set out to study the security implications of a recent vulnerability, prototype pollution, in JavaScript runtime environments.

Prototype pollution is a vulnerability affecting the JavaScript language~\cite{arteau2018prototype}. JavaScript's prototype-based inheritance allows an object to inherit properties from its ancestors via the prototype chain. When accessing a property not present on the object, the prototype chain will be queried for that property instead. Unless explicitly changed, this chain connects all objects to a common root prototype. Pollution can occur when an attacker-controlled value is used to navigate an object's structure. Since each object has a runtime accessible reference to its prototype, the attacker may be able to pick that reference and add a new property. By doing this, the attacker can cause a change in behavior in another part of the application.

The security implications of prototype pollution depend on the presence of otherwise benign pieces of code (gadgets) that inadvertently read attacker-controlled properties  from the root prototype to execute sensitive operations, e.g., arbitrary code. Gadgets in JavaScript runtime environments are particularly dangerous because they are shared by all applications, thus increasing the attack surface.

The vast majority of prior works focus on the detection of prototype pollution by static analysis \cite{kim2021dapp,Xiao21,Li21,Li22,ShcherbakovBS23}, while the existence of gadgets remains largely unexplored \cite{Kang22,ShcherbakovBS23,shcherbakov2024unveiling,liu2024undefined}. This work is inspired by the recent pioneering of work of Shcherbakov et al. \cite{ShcherbakovBS23}, which uses static taint analysis for three Node.js APIs to find (combinations of) three gadgets, dubbed \emph{universal gadgets}, leading to arbitrary code execution. Our thesis is that dynamic analysis should be preferable for identifying universal gadgets for these reasons: (a) the sources of the analysis pertain to accesses of properties from the prototype, which is hard to identify statically; (b) the highly-dynamic nature of JavaScript poses significant challenges for static analysis, resulting in low precision and recall,  and high manual effort \cite{ShcherbakovBS23}; (c) realistic gadgets should trigger in common use cases of API usages, which is best captured by the comprehensive test suite of runtime environments.

To address these challenges, we design, implement, and evaluate a semi-automated pipeline, \tool{}, to comprehensively and systematically detect universal gadgets in V8-based JavaScript runtimes, Node.js and Deno.  Deno is a particularly interesting target because it is proposed as a security-first runtime to counter the  shortcomings of Node.js. Specifically, \tool{}  customizes Deno, Node.js, and the V8 engine to implement a  lightweight dynamic taint analysis for automatically identifying gadget candidates, which we validate manually to derive proof-of-concept exploits.  Driven by the test suite of a runtime environment,  \tool{}  detects property accesses from an object's prototype, it injects a taint value, and monitors the execution to identify the effects of the taint value on security-sensitive sinks and unexpected terminations. Moreover, \tool{}  supports processing and representation of gadget candidates in SARIF format \cite{sarif} for visualization  to facilitate the manual analysis.

We use \tool{} in a comprehensive study of Node.js and Deno to identify universal gadgets pertaining to a range of vulnerabilities, including arbitrary code execution, server-side request forgery, privilege escalation, cryptographic downgrade,  and more. After processing, \tool{} automatically identifies 301 and 418 gadget candidates in Node.js and Deno, respectively. We manually verified the gadget candidates to find 56 universal gadgets in Node.js and 67 universal gadgets in Deno for a total of 28 person-hours. 
We further compare \tool{} with Silent Spring~\cite{ShcherbakovBS23}, showing that it provides increased precision and recall, while reporting less gadget candidates for manual analysis. To support further research on the topic, we make available publicly both \tool{}~\cite{artifact} and the gadgets~\cite{gadgetsKTH}.

We have responsibly disclosed our findings to the Node.js and Deno development teams. Both acknowledged our report but neither considers them within their current thread model. Node.js suggested a public discussion with their developers' community on the dangers of gadgets.

In light of these results, we systematize, for the first time, existing mitigations for prototype pollution and gadgets in terms of development guidelines. We then collect a list of applications with end-to-end exploits pertaining to prototype pollution, and revisit the fixes through the lens of our guidelines. Through this exercise, we also identify existing issues, including one high-severity CVE-2023-31414 leading to remote code execution, which was due to incorrectly fixing a gadget.

Our contributions can be summarized as follows:
\begin{itemize}
  \item We design and implement a semi-automated pipeline, \tool{}, to systematically detect universal gadgets in JavaScript runtimes (Section~\ref{sec:methodology}). 
  \item We conduct a comprehensive analysis of Node.js and Deno to find 123 universal gadgets subject to a range of vulnerabilities (Section~\ref{sec:eval}).
  \item We systematize  existing mitigations against prototype pollution and gadgets, and outline directions for future work, including  an in-depth case study leading to RCE (Section~\ref{sec:defense}).
\end{itemize}

\section{Technical Background}
\label{sec:background}

In this section, we overview the life cycle of exploits pertaining to prototype pollution vulnerabilities, and discuss the JavaScript runtime of interest and the threat model.

\begin{figure}[t]
\centering
\begin{lstlisting}[caption={Example of prototype pollution and gadget.},label={lst:background-gadget}]
const users = { };
router.post("/:uid", (req, res) => {
  users[req.uid][req.key] = req.value;
  exec("echo 'A value was stored at' $(date)");
  res.status(200).send();
});
function exec(cmd, opts) {
  opts = opts || {};
  const shell = opts.shell || "/bin/sh";
  op_spawn(`${shell} -c '${sanitize(cmd)}'`);
}
\end{lstlisting}
\end{figure}

\subsection{Prototype Pollution and Gadgets}

Prototype pollution is a vulnerability that occurs in prototype-based languages like JavaScript \cite{arteau2018prototype}. An attacker manipulates a program's prototype-based inheritance, leading to runtime modification of objects and potentially causing otherwise benign code sequences, called gadgets, to execute dangerous operations.  End-to-end exploitation of  gadgets based in prototype pollution requires two steps. The prototype must be polluted first, for example when processing untrusted user data incorrectly, and then the gadget must be triggered.

To illustrate the vulnerability, Listing \ref{lst:background-gadget} shows an artificial server application which provides an in-memory key-value store for its users, logging every request to standard output. It is vulnerable to prototype pollution and uses function \verb|exec| as a gadget. \verb|exec| (line 7-11) is an otherwise benign runtime-provided function to execute a command. It accepts the command to execute as a string and an optional object \verb|opts| to configure the shell in which to execute the command.

A request at \verb|vuln.com/uid?key=value| causes the server to invoke the handler on line 2-6.  It extracts the user ID and the key-value pair from the URL and stores it in memory (line 3). It then logs the time of the request  (line 4) and responds with a \verb|200| status code (line 5).

An attacker can use this handler to perform prototype pollution. The malicious request \verb|vuln.com/__proto__?shell=node -e '...'| will add the property \verb|shell| with the value \verb|"node -e '...';"| to the root object prototype on line 3. This happens because the request instantiates the statement on line 3 as \verb|users["__proto__"]["shell"] = "node -e '...';"|. In particular, \verb|users["__proto__"]| gives a reference to \verb|Object.prototype| which is then extended with the property \verb|shell|.

The attacker can capitalize on the pollution of the \verb|shell| property to turn the benign call to \verb|exec| into a remote code execution gadget. In particular, because the application provides no options on line 4, line 8 assigns to \verb|opts| an empty JavaScript object. When evaluating the expression \verb|opts.shell| on line 9, the \verb|shell| property, missing from \verb|opts|, will be looked up in the prototype chain where it exists because of the pollution. Thus, \verb|opts.shell| evaluates to \verb|"node -e='...';"| and is used instead of the default \verb|"/bin/sh"| to evaluate arbitrary JavaScript code.

\subsection{JavaScript Runtimes: Node.js and Deno}

In this work, we  study universal gadgets in JavaScript runtime environments. 
Two such runtime environments are Node.js and Deno. Both are open source software projects built on top of the V8 JavaScript engine from Chromium. Node.js is a popular JavaScript runtime \cite{Nodejs} written in C++, commonly used for server application development. Deno was created in response to Node.js with a focus on security \cite{Deno}. It is written in Rust and uses TypeScript. The native (C++/Rust) parts of these runtimes are what provides access to system resources and common functionality such as buffers and cryptography libraries. In this work we focus on these runtimes because of their popularity and shared JavaScript engine.

Deno's focus on security is interesting for our work because it adds guardrails for both pollution and gadgets. On the pollution side, Deno removed the \verb|__proto__| property, rendering the  attack described on Listing \ref{lst:background-gadget} infeasible. However, prototype pollution is still possible through, e.g., object merge functions, a common source of prototype pollution. On the gadget side, Deno has a permission system to reduce access to system resources and by extension the impact of gadgets. However, we observe that the presence of a gadget implies some access to the corresponding resource must have been granted to the application, thus allowing exploits nonetheless.

\subsection{Threat Model}
Our threat model focuses on server-side JavaScript/TypeScript applications running on either Node.js or Deno. We assume the application is vulnerable to prototype pollution, either directly or through third-party code. Our aim is to find exploitable universal gadgets present in the JavaScript runtime for the purpose of one of (directly or indirectly):

\begin{itemize}
  \item Arbitrary Code/Command Execution (ACE). Gadgets that allow an attacker to execute arbitrary JavaScript code or start an arbitrary command.
  \item Server Side Request Forgery (SSRF). Gadgets that allow an attacker to make arbitrary network requests.
  \item Privilege Escalation. Gadgets that allow an attacker to perform an action their normal privileges do not allow.
  \item Cryptographic Downgrade. Gadgets that downgrade the cryptography used by the application to be weaker.
  \item Path Traversal. Gadgets that allow the attacker to manipulate the path of file system operations.
  \item Unauthorized Modifications. Gadgets that allow the attacker to trigger modifications that should not happen as a result of normal operation.
  \item Log Pollution. Gadgets that change or control the contents of program logs.
  \item Denial of Service (DoS). Gadgets that deny access to the application.
\end{itemize}

We posit that many applications use some of these APIs in practice because of the importance of the functionality they provide. Furthermore, we assume that the runtime's own test suite contains a representative sample of ways to use the APIs. As a direct consequence, the presence of a gadget in a runtime implies vulnerabilities in real-world applications.

\section{Overview}
\label{sec:overview}

\begin{figure}[t]
	\centering
	\includegraphics[width=\linewidth]{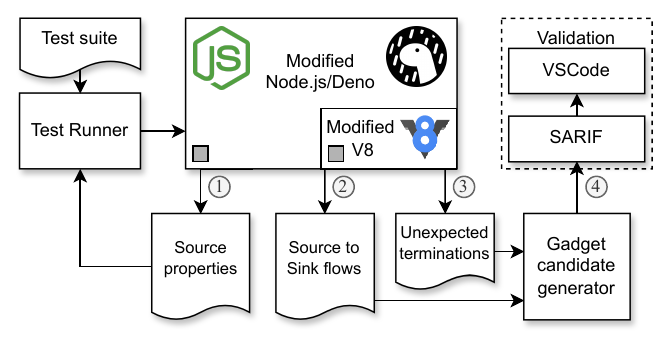}
	\caption{Architecture and workflow of \tool{}.}
	\label{fig:analysis}
\end{figure}

At a high level we develop  a semi-automated dynamic analysis pipeline, \tool{}, for finding gadgets in runtime environments, as depicted in Figure \ref{fig:analysis}. 
To achieve this goal, \tool{} operates in three automated steps and one manual step. Driven by the
runtime's test suite, the first step identifies candidate properties for prototype pollution by detecting undefined property accesses. In the second and third step, \tool{} uses these candidate properties to simulate pollution and detect reachability of dangerous sinks and unexpected termination, respectively.  These steps also rely on the runtime's test suite and generate output for gadget identification. The final step consists in manually verifying the results of the second step, after preprocessing, using visualization of SARIF files in IDEs, and generating proof-of-concept exploits. 

Listing \ref{lst:fetch-source}  shows a universal gadget in Deno, which we will use to illustrate the workflow of \tool{}  along with the different challenges we have to tackle. Consider an application that uses the runtime API \verb|fetch|, defined in Listing \ref{lst:fetch-source}, to fetch user details from another service, for a given trusted user identifier \verb|uid|. The application will eventually execute the command \verb|fetch("https://192.168.3.14/users/"+uid)|  to safely retrieve user information. Given the assumption that the application is vulnerable to prototype pollution,  our goal is to find out how we can use prototype pollution to turn this seemingly benign request into a malicious gadget.

\tightpar{Step 1: Collecting source properties} A key requirement is to find properties that influence the behavior of a runtime API. 
These properties must not be defined so that they are looked up in the prototype chain and a polluted value is used instead. Hence, \tool{}  needs to determine which undefined property accesses happen as a result of normal usage of a target runtime API. This is achieved by observing the runtime behavior of code and taking note of undefined property accesses. Moreover, \tool{}  uses the runtime environment's test suite as a representative sample of normal usage of the API.

For the \verb|fetch| API in Listing \ref{lst:fetch-source}, \tool{} runs Deno's test suite to collect a list of undefined properties that includes \verb|method| (line 3) and \verb|signal| (line 9). This leads us to our first challenge of automatically identifying undefined property accesses driven by the test suite of runtime APIs, which we discuss in Section \ref{sec:methodology:undefined-properties}. 

\tightpar{Step 2: Identifying source-to-sink flows} \tool{} uses the list of undefined property accesses from the previous step as sources for further analysis. To determine if a property is used for a purpose that is exploitable, \tool{} implements a lightweight taint analysis that identifies the reachability of values of  polluted properties into dangerous sinks. Driven by the test suite, it pollutes the undefined properties with taint values and checks whether these values affect the native (C++/Rust) code of the runtime environment, which conservatively represents security-relevant sinks. 

The function call to \verb|op_fetch| in Listing \ref{lst:fetch-source} (line 13) executes Deno's native networking implementation for \verb|fetch|.
To determine if a polluted value can reach \verb|op_fetch|, \tool{} simulates prototype pollution and  detects the polluted property value in the call to \verb|op_fetch|.  For the property \verb|method|, \tool{} pollutes the property with a taint value and runs the corresponding test case, while intercepting every call to \verb|op_fetch| and checking all arguments for the presence of the taint value used for pollution.
Indeed, given the list of properties for \verb|fetch|, \tool{} finds that the property \verb|method| reaches the sink \verb|op_fetch| on line 13. This leads us to our second challenge of automatically identifying flows from undefined properties to sinks, which we discuss in Section \ref{sec:methodology:dangerous-sinks}. 

\begin{figure}[t]
\centering
\begin{lstlisting}[caption={Simplified Deno fetch implementation.},label={lst:fetch-source}]
class Request {
  constructor(input, init = {}) {
    this.method = init.method || "GET";
    // ...
  }
}
function fetch(input, init = {}) {
  const request = new Request(input, init);
  const promise = mainFetch(request, false, request.signal);
  //...
}
async function mainFetch(req, recursive, terminator) {
  const res = op_fetch(req.method, /*...*/);
  terminator[abortSignal.add]();
  //...
}
\end{lstlisting}
\end{figure}

\tightpar{Step 3: Unexpected termination}
If normal usage of a runtime API (as represented by the test suite) does not result in a crash but the pollution of an undefined property does cause the API to crash, it implies that an attacker can use the API to cause Denial of Service (DoS) attacks. Similarly to Step 2, \tool{} leverages the runtime's test suite to detect DoS attacks pertaining to prototype pollution. When polluting the property \verb|signal| on line 9, \tool{} causes the \verb|fetch| API to crash due to a type error on line 14. This leads us to our third challenge of automatically identifying fatal crashes that cause DoS attacks on applications that use the APIs under pollution, which we discuss in Section \ref{sec:methodology:unexpected-termination}. 

\tightpar{Step 4: Manual validation} The previous automated steps yield a list of potential sinks and unexpected program crashes pertaining to pollution of undefined properties. These results do not necessarily imply that a runtime API is exploitable, but require manual validation. To aid the security analyst, \tool{} supports processing (e.g., removal of duplicates from different test cases) and representation of results in SARIF format for visualization within an IDE.

In our example, the  SARIF file contains two results, called gadget candidates, for the \verb|fetch| API: One for property \verb|method| reaching the sink \verb|op_fetch| and one for property \verb|signal| resulting in a program crash. The manual analysis of \verb|method| reveals that an attacker can override the default HTTP method from \verb|GET| at wish, revealing a true gadget. For instance, they can pollute \verb|method| with value \verb|DELETE|, thus causing the command \verb|fetch("https://192.168.3.14/users/"+uid)| to delete user records (in Section \ref{sec:eval} we extend this attack to full Server Side Request Forgery). The analysis of the program crash due to \verb|signal| reveals an attacker can perform a DoS attack, thus denying users of access to data. In Section \ref{sec:methodology:manual-verification} we discuss this final challenge of effectively validating  gadget candidates.

\section{System Design and Implementation}
\label{sec:methodology}

We design \tool{} to overcome the challenges outlined in Section \ref{sec:overview}. In line with the architecture and workflow of Figure \ref{fig:analysis}, this section describes and motivates our design and explains how it supports comprehensive analysis of JavaScript runtime environments for finding gadgets. First, we discuss source properties and detail our approach to capturing them exhaustively. Second, we show how to achieve comprehensive coverage for sinks into native runtime code and how to identify source-to-sink flows by our lightweight taint analysis. Third, we discuss unexpected termination and how to detect fatal terminations leading to DoS attacks. Finally, we discuss the process of preprocessing and manually validating results, as well as the current limitations of \tool{}. 

Along with the discussion of the design we also describe the implementation of \tool{}, which we implement against Node.js \verb|v21.0.0| and Deno \verb|v1.37.2|. These are the most recent versions of the respective runtimes that share a common V8 engine version, namely \verb|v11.8.172.17|.

\subsection{Source Properties}
\label{sec:methodology:undefined-properties}

In this work we consider undefined property accesses as \emph{sources}. At a high level, an undefined property access happens when code tries to read a property that is not one of the object's own properties. There are many ways in which this can happen in JavaScript, including \verb|obj.prop| as seen on line 3 of Listing~\ref{lst:fetch-source}, computed property names such as \verb|obj[str_var]|, array-indexed properties such as \verb|obj[1]|, for-in loops, and various syntactic sugar forms such as destructuring assignment. These features pose significant challenges for static analysis approaches \cite{ShcherbakovBS23}, leading to both false positives (due to conservatively computing undefined properties) and false negatives (due to computed property names).

To ensure we comprehensively capture all undefined property accesses we modify the V8 runtime to trap on property accesses that are looked up but not present in the root object's prototype object. This conservatively covers all property accesses that may be influenced by prototype pollution, excluding pollutions with other side effects (i.e. existing prototype properties) and circumstantial pollutions of specific types.

Because gadgets are pre-existing runtime API function calls in application code, we are interested in undefined property accesses that happen as a result of normal API usage. Thus, we leverage the runtime's test suite as a proxy of real API usage and capture all undefined property access that occur during test execution. We store the observed property names on a per-test basis for use in the next  steps.

For our example of Section \ref{sec:overview} this step yields  95 properties for \verb|fetch| from the \verb|fetch_test.ts| test suite in Deno.

\tightpar{Implementation}
To intercept all property accesses, we modify the code of \verb|Runtime::GetObjectProperty| and \verb|LoadIC::Load| methods, which look up the property name in an object's prototype chain to read a property value. If the property is not found in the chain we log the access attempt.

However, V8 implements optimizations to avoid slow calls to these methods when the property name can be easily determined, as in  \verb|obj.prop|. Thus, we deoptimize the inline caches~\cite{v8-inline-caches} and remove the bytecode handlers in the methods \verb|AccessorAssembler::LoadIC_NoFeedback| for named properties and \verb|AccessorAssembler::KeyedLoadIC| for array-indexed properties. This allows us to trap on every  property  access, albeit with some performance degradation.

We also implement a separate file logger to dump the results of our tests and extend the globals object with the \verb|log| function. This enables our modifications in the test suite to use the same logs for dumping call stacks as described later in this section. The changes to V8 are limited to 8 files and modify 233 lines of code in total.

\subsubsection{Simulating Pollution}
\label{sec:methodology:simulating-pollution}

Given the names of undefined properties that are accessed for a test, we want to simulate pollution of these properties to observe how it affects the behavior of the runtime. To this end we extend the test runners to automatically modify test files by injecting a code snippet that simulates prototype pollution.

To maximize effectiveness, the polluting snippet is injected at the top of the test file. This ensures the entire test execution is affected by the pollution. In comparison to injection using preloaded modules (e.g. through \verb|--require| or \verb|--module| in Node.js) this avoids affecting irrelevant accesses that happen before the test is started.

We use this prototype pollution simulation in the next two steps. In particular, if $N$ unique undefined property accesses were detected for a test, 
we run for both the second and third stage of  \tool{}  with $N$ different instances of that test, each with a different property polluted.

For our  example this means the \verb|fetch_test.ts| test file in Deno is dynamically updated on the fly with a snippet that pollutes one of the 95 detected properties at a time.

\tightpar{Implementation}
We use two types for the injected values: strings and objects. To assign the property we use \verb|Object.defineProperty| to add gettable (and settable) value. This allows us to output a stack trace for all accesses to that property. Additionally, we utilize this getter to return a unique identifier (incremental number) for every access so that we can match sources and sinks by the tainted value. Listing \ref{lst:simulating-pollution} shows the injected snippet for string values, while the snippet for object values is similar \cite{artifact}.

One of the values we use is a hexadecimal string so that it can be converted into a number, if needed. To support code that expects \verb|Object| as the type for polluted values, we inject objects built based on JavaScript \verb|Proxy|. These tainted values emulate the reading of arbitrary properties via \verb|ProxyHandler|, access to an iterator to support for-of loop against this object, and conversion to primitive types. Each of these access methods also produces a tainted value to propagate the taint mark.

\subsection{Source-to-Sink Flows}
\label{sec:methodology:dangerous-sinks}

We consider function calls where JavaScript executions flow into the runtime's native code as \emph{sinks}. To be able to exhaustively cover such sinks we study the ECMAScript standard~\cite{ECMA335} to determine function calls that flow into V8 as well as the runtime's development documentation to understand where such flows occur for the runtime's native modules.

For V8, we find that functions such as \verb|eval| and \verb|new Function()| are the sinks that create a function at runtime from their string  arguments. 
In particular, both functions create and subsequently execute JavaScript code. Thus, if a polluted value is used as (part of the) input to these functions, an attacker can potentially execute arbitrary  code.

For Node.js, based on its contributor documentation~\cite{nodejs-fast-api} and source code, we identified internal APIs that interoperate with the C++ implementation from JavaScript: \emph{linked bindings} and \emph{internal bindings}. After conducting tests, we confirmed that \emph{linked bindings} are intended for developers to extend Node.js with additional C++ bindings, and this method is not used for Node.js runtime APIs.
Consequently, we determined that \emph{internal bindings} comprehensively cover all data flows from JavaScript to the C++ part of Node.js and are implemented in a single JavaScript file: \verb|lib/internal/bootstrap/realm.js|.

For Deno, similar to Node.js, we identify \emph{bindings} as the only bridge between JavaScript and Rust. This is based on the contributor documentation for \verb|#[op]| and \verb|#[op2]| Rust attributes used throughout the Deno code base. As a result we identify a single template file written in JavaScript in the \verb|deno_core| codebase that comprehensively covers all flows from JavaScript to Rust: \verb|core/runtime/bindings.js|.

When the sink receives a tainted value as one of its arguments, it logs information about the sink being reached. This  includes the sink name, call stack, tainted value with an identifier for source matching, and the access path if the tainted value is detected in a nested property of the argument. 

For the running example of Section \ref{sec:overview} this step yields only one result in Deno, namely that of pollution of the \verb|method| property into the \verb|op_fetch| binding.

\tightpar{Implementation}
To capture flows involved in creating functions at runtime, we modified the method \verb|Compiler::GetFunctionFromEval()|. This method generates a function from a string passed into its first argument. Public APIs such as \verb|eval| and \verb|new Function()| use this method. We test the value of the first argument, and if it contains our tainted mark as a substring, we log the argument's value along with a record that this sink was triggered.

To capture the flows via binding code  we implement a wrapping layer that we apply to all bindings for both runtimes. This wrapper recursively replaces all functions on a JavaScript object with a new function that inspects the arguments for tainted values, calls the original function, and returns its result. If a tainted value is detected we log the sink name, the argument index, the current stack trace, and (if applicable) the path to the tainted value for objects (e.g. \verb|x| if the value of property \verb|o.x| was tainted). This wrapper consists of approximately 380 lines of JavaScript code and is used in both \verb|realm.js| and \verb|bindings.js| for Node.js and Deno respectively.

\subsection{Unexpected Termination}
\label{sec:methodology:unexpected-termination}

Besides dangerous sinks we are also interested in pollutions that result in unexpected or non-termination of the program, indicating potential DoS attack. We focus on fatal crashes that JavaScript code cannot catch and thus terminates the application immediately. Because crashes may happen with no tainted value reaching a sink, we perform this evaluation separately.  \tool{} can also  detect non-fatal crashes (catchable in JavaScript), which we do not include in our results. 

To comprehensively cover unexpected termination as a result of pollution, we monitor all test executions and look for processes that exit with a non-zero exit code. If a non-zero exit code is detected we evaluate the stdout and stderr of the process to filter out expected failures such as test failures in order to report only unexpected errors such as segfaults/panics, Out Of Memory (OOM), and timeouts.

To avoid reporting crashes that may happen as a result of our runtime modifications, we perform this analysis on the original runtimes. This works because this stage relies exclusively on externally available information, namely the previously-obtained list of undefined property accesses.

For the running example of Section \ref{sec:overview} this step yields only one result in Deno, namely that of pollution of the \verb|signal| property leading to an unexpected \verb|TypeError|.

\tightpar{Implementation}
To perform this part of the analysis, we re-use the test runner that modifies test files with prototype pollution and instruct it to use the unmodified version of the runtime. We extend the test runner to examine the exit code and output (\verb|stdout| and \verb|stderr|) for each test it runs. In particular, if the exit code is nonzero, it will check if the output matches an expected error (e.g. a test failed) and if it does not, log the polluted property name and process output.

\subsection{Manual Validation}
\label{sec:methodology:manual-verification}

To  effectively validate and create proof-of-concept exploits from the results of Section \ref{sec:methodology:dangerous-sinks} and Section \ref{sec:methodology:unexpected-termination}, we produce a SARIF file with all necessary information for manual validation. The SARIF file format, in combination with a SARIF file viewer, provides a convenient way for an analyst to interactively view results and browse relevant code locations.

We preprocess the output of stages 2 and 3 to obtain a \emph{gadget candidate} for each unique detected sink or unexpected termination. For a reached sink, this is determined by the property name and the stack trace for the sink call or the stack trace for the polluted property access. For unexpected termination, this is determined by the termination output.

For each gadget candidate, we include all relevant information for validation and creation of a proof of concept. For detected sinks the gadget candidate is presented as a triple consisting of the polluted property name as well as the API and sink represented by the stack trace for the source and sink (SARIF viewers allow for interactively browsing the stack). We also provide the value observed at the sink which helps the analyst understand if the runtime manipulates the polluted value. For unexpected terminations, we are limited to providing the program output after the crash, but additionally we provide the name of the polluted property as well as the test file that crashed.

While each result represents only a single polluted property, if multiple properties affect the same API and sink these results will be co-located in the generated SARIF file. This allows the analyst to combine multiple properties in a proof of concept. Thus, in contrast to a gadget candidate, a \emph{gadget} is a triple consisting of the set of properties,  API and sink.
We remark that \tool{} only detects that a value reaches the sink but not the intended type or structure of that value. The analyst has to analyze the  API documentation and code to understand what values to use in the proof-of-concept exploit.

For the running example of Section \ref{sec:overview}, the SARIF file contains two entries, one for the detected flow from the property \verb|method| to the sink \verb|op_fetch| and one for the unexpected error as a result of polluting the property \verb|signal|.

\tightpar{Implementation}
We generate the SARIF file from the logs of the second and third stages. For  the second stage we look for sinks where a tainted value was observed and the corresponding source (property access for that exact value). As a result any source that does not reach a sink is automatically discarded. If  no source can be found for a taint value at a sink (e.g. due to modifications to the value), it is reported to the analyst separately. For the third stage we report any test run resulting in a non-zero exit code with a stderr message other than a test failure, excluding tests that failed in the initial run.

\subsection{Limitations}
\label{sec:limitations}

\tightpar{Full-fledged taint tracking}
Our lightweight taint analysis favours performance. This can be seen as a limitation with respect to manual validation because the complete flow from source to sink is not readily available. In practice, we find that the runtime code is relatively simple for most cases,  and the flow from source to sink can be identified quickly. Secondly, our lightweight taint tracking may miss flows from sources to sinks in the event that the taint value is removed in certain operation (e.g. splice). Again, we observe that most runtime code does not perform modifications on values beyond simple transformations such as converting a string to uppercase.

\tightpar{Polluted types}
The pollution simulation only pollutes using strings and objects. We could additionally cover numbers and arrays for pollutions (booleans cannot be taint tracked with our approach). This would only find flows where an explicit type check prevents the tainted value from reaching a sink.
Besides polluting with different types, techniques such as concolic execution \cite{liu2024undefined,steffens2021understanding} could be used to improve coverage too.

\tightpar{Gadget chains}
In contrast to works on gadget detection in libraries and frameworks \cite{liu2024undefined,shcherbakov2024unveiling}, \tool{} cannot find gadget chains where one pollution enables another.  This is because \tool{} pollutes only a single property at the time. Running an analysis where multiple properties are polluted at the same time is possible in theory, but infeasible in practice due to the number of possible combinations of properties.

\tightpar{Binding coverage}
For Node.js we are unable to cover 25 bindings because they exist at a property that is not configurable or not writable, thus preventing us from wrapping them. We evaluated these functions and find them to have little security relevance. For Deno we were unable to wrap 4 bindings, all async, because they do not take any arguments. Such sinks are not interesting for our analysis so we consider this a non-issue.

\tightpar{Test suite limitations}
Our approach relies on the comprehensiveness of the runtime's test suite. We are thus limited in our analysis by the coverage of the source code by the test suite. We evaluate the coverage statistics and find 95.8\% and 91.4\% function coverage in Node.js and the Deno standard library respectively. These percentages give confidence in the comprehensiveness of our analysis.

\section{Evaluation}
\label{sec:eval}

This section describes the results of our comprehensive evaluation on Node.js and Deno, answering the research questions:

\begin{itemize}
  \item \textbf{RQ1:} How can we effectively identify exploitable universal gadgets in the Node.js and Deno runtimes?
  \item \textbf{RQ2:} How does \tool{} compare to Silent Spring?
  \item \textbf{RQ3:} What is the performance overhead of our taint-enhanced runtimes as compared to the original runtimes? How to  empirically validate transparency of our taint-enhanced runtime with respect to the original runtimes?
\end{itemize}

\tightpar{Experimental setup}
We conduct our experiments on an AMD EPYC 7742 64-Core 2.25 GHz server with 512 GB of RAM. To optimize server resource utilization, we execute tests in parallel.
We utilize a modified test runner script that runs test files in parallel with a 20 second timeout per test file. For Node.js we adopt the existing \verb|test.py| runner, for Deno we write a custom runner that invokes \verb|deno test|.

\subsection{Universal Gadgets in Node.js and Deno}

We demonstrate the effectiveness of \tool{} through the number of detected gadgets in light of the number of outputs for intermediate analysis steps. 

\tightpar{Analysis of Node.js}
The target of our analysis of Node.js is the standard library built into the Node.js binary.
The first step of our analysis produced 509,481 unique test-property combinations for 3,782 test files.
The second and third steps of our analysis found 22,860,092 sinks reached, 9,743 segfaults, and 6 tests that timeout. Preprocessing of results reduced the number of sink-source pairs to 13,029 unique pairs and segfaults to 13 (no reduction in test timeouts). Furthermore, we excluded source-sink pairs that could only lead to Denial of Service: 
11,730 sinks related to infrastructure code such as type checking, internal utils, asynchronous call wrappers, exception and error message builders; 
120 in \verb|buffer.byteLengthUtf8|; 
258 in \verb|messaging.postMessage|, which sends messages between workers; 
and 101 in the \verb|buffer| parameter in \verb|fs.read| which is used for output of the sink call. 
After filtering, there are 820 gadget candidates out of which we confirmed 56 to be exploitable. The manual verification process required 31 person hours. 

\tightpar{Analysis of Deno}
Our analysis of the Deno runtime covers the core API (accessible by \verb|Deno|), the Node.js compatibility module, and the Deno standard library. We ran our pipeline on each separately, but accounted for duplicates when aggregating the results, which we report here.

The first step of our analysis produced 21,786 unique test-property combinations for 596 test files. The second and third steps of our analysis found 13,519 sinks reached, 1 panic, and 139 tests that timeout. Preprocessing of results reduced the number sink-source pairs to 399 unique pairs, 18 tests that timeout, and no reduction in panics. As a result, we obtained 418 gadget candidates out of which we confirmed 67 to be exploitable. The manual validation took 15 person hours. 

\tightpar{Node.js vs Deno}
We observe quite a large difference in numbers when comparing Node.js to Deno. First, Node.js produces significantly more results. One reason for this is that Node.js has a larger test suite (both in terms of test files and test cases).
Despite Deno's security focus, we find similar number of exploitable gadgets. One reason for this is that Deno has a larger API surface. Another is that prior work on gadgets has resulted in some protections being implemented in Node.js, in fact some of the gadgets we find in Deno were previously identified and addressed in Node.js.

\tightpar{Result classification}
We categorize our universal gadgets by the strongest exploit they can be used for. If multiple properties can be combined to achieve a stronger exploit, we consider only the combination and not the weaker exploits pertaining to a subset of properties. Table \ref{table:gadgets-overview} shows the aggregate number of gadgets per exploit category.

We omit gadgets without a security impact or that only cause a JavaScript exception (they have limited impact since applications can catch such exceptions). We include gadgets that presume an existing vulnerability (e.g. to write a file on the systems) and call these \emph{second order} gadgets.

\begin{table}
  \footnotesize
  \centering
  \begin{tabular}{ |r|c|c| } 
    \hline
    \textbf{Attack Type}              & \textbf{Node.js} & \textbf{Deno} \\
    \hline
    Arbitrary Code/Command Execution  & 14               & 5             \\ \cline{2-3}
    Server Side Request Forgery       & 6                & 3             \\ \cline{2-3}
    Privilege Escalation              & 7                & 24            \\ \cline{2-3}
    Cryptographic Downgrade           & 2                & 0             \\ \cline{2-3}
    Path Traversal                    & 3                & 10            \\ \cline{2-3}
    Unauthorized Modifications        & 0                & 10            \\ \cline{2-3}
    Log Pollution                     & 0                & 1             \\ \cline{2-3}
    Panic/Segfault                    & 12               & 1             \\ \cline{2-3}
    Out of Memory                     & 0                & 3             \\ \cline{2-3}
    Infinite Loop                     & 0                & 2             \\ \cline{2-3}
    Second Order                      & 12               & 8             \\ 
    \hline  
    Total                             & 56               & 67            \\
    \hline
  \end{tabular}
  \caption{Number of gadgets found by type per runtime.} 
  \label{table:gadgets-overview}
\end{table}

\tightpar{New detected gadgets} 
We highlight 4 gadgets here and refer to Table \ref{tab:nodejs-gadgets} and Table \ref{tab:deno-gadgets} in Appendix, and code artifact \cite{artifact} for the complete list of gadgets.

Listing \ref{lst:deno-gadget-fetch} shows a proof of concept (PoC) of the \verb|fetch| gadget from Section \ref{sec:overview}. In addition to the property \verb|method|, polluting the properties \verb|body| and \verb|headers|  allows attackers to control all aspects of the request to the application-specific URL. Moreover, due to the way Deno's \verb|fetch| implementation stores request URLs internally, the pollution of property \verb|0| allows the attacker to override the URL and achieve SSRF. This gadget  transforms a simple benign-looking request like \verb|fetch("http://example.com")| into a completely unrelated HTTP request.

\begin{lstlisting}[caption={PoC of \textit{fetch} gadget (Deno).},label={lst:deno-gadget-fetch}]
// send a POST request to http://fake.com
///////////////////////////////////////////////
// PROTOTYPE POLLUTION:
Object.prototype[0] = 'http://fake.com'
Object.prototype.method = 'POST'
Object.prototype.body = '{"pwned":"yes"}'
Object.prototype.headers = {"content-type":"application/json"}
///////////////////////////////////////////////
// GADGET:
fetch('http://example.com')
\end{lstlisting}

Similarly, we found that the \verb|fetch| API of Node.js can also exploited to achieve SSRF attacks. In addition to controlling \verb|method| and \verb|body|, an attacker is able to pollute \verb|socketPath| to redirect HTTP requests to a local socket rather than the specified URL. This gadget can be exploited to target local daemons, such as Docker.

Another universal gadget in Deno allows for path traversal on temporary files. Polluting \verb|dir| allows an attacker to control where \verb|Deno.makeTempDir| and \verb|Deno.makeTempFile| create temporary file system entries. Even if \verb|dir| is specified by the application, \verb|prefix| still allows for path traversal by using a string like \verb|../| as a prefix (prior to Deno v1.41.1). Depending on how the temporary file is used, this gadget can be a setup for a stronger attack.

We also identify two new Arbitrary Code Execution (ACE) gadgets in Node.js, located in the commonly used \verb|require| and \verb|import| functions. The gadget in \verb|require| has been fixed as of Node.js v18.19.0. 
We detail this gadget and its fix in Section~\ref{sec:case-studies}. The gadget associated with \verb|import|, shown in Listing \ref{lst:node-gadget-import}, can be exploited by polluting the \verb|source| property with JavaScript code and invoking the \verb|import| function on any \verb|.mjs| file. This causes the code from the property to be evaluated.

\begin{lstlisting}[caption={PoC of \textit{import} gadget (Node.js).},label={lst:node-gadget-import}]
///////////////////////////////////////////////
// PROTOTYPE POLLUTION:
Object.prototype.source ='console.log("PWNED")'
///////////////////////////////////////////////
// GADGET:
import('./any_file.mjs')
\end{lstlisting}

\begin{table}
  \scriptsize
  \centering
  \begin{tabular}{ | r | c | c | c | c | c | c | c | }
    \hline
    \multirow{2}{*}{API}      & \multirow{2}{*}{GT} & \multicolumn{3}{c}{Silent Spring} & \multicolumn{3}{|c|}{\tool{}} \\\cline{3-8}
                              &                     & GC       & TP/FP     & FN         & GC       & TP/FP   & FN       \\\hline
    \verb|cp.exec|            & 2                   & 20       & 1/19      & 1          & 3        & 2/1     & 0        \\       
    \verb|cp.execFile|        & 1                   & 16       & 0/16      & 1          & 2        & 1/1     & 0        \\       
    \verb|cp.execFileSync|    & 4                   & 21       & 3/18      & 1          & 7        & 4/3     & 0        \\       
    \verb|cp.execSync|        & 4                   & 13       & 3/10      & 1          & 7        & 4/3     & 0        \\       
    \verb|cp.fork|            & 2                   & 25       & 1/24      & 1          & 6        & 2/4     & 0        \\       
    \verb|cp.spawn|           & 3                   & 14       & 2/12      & 1          & 5        & 3/2     & 0        \\       
    \verb|cp.spawnSync|       & 4                   & 11       & 3/8       & 1          & 7        & 4/3     & 0        \\       
    \verb|import|             & 1                   & 0        & 0/0       & 1          & 5        & 1/4     & 0        \\       
    \verb|require|            & 3                   & 19       & 2/17      & 1          & 4        & 1/3     & 2        \\       
    \verb|vm.compileFunction| & 1                   & 4        & 1/3       & 0          & 5        & 0/5     & 1        \\\hline 
    Total                     & 25                  & 143      & 16/127    & 9          & 51       & 22/29   & 3        \\\hline
  \end{tabular}
  \caption{Silent Spring vs \tool{} on Node.js v16.13.1 with properties used in Silent Spring gadgets as ground truth.}
  \label{tab:comparison-16}
\end{table}

\subsection{GHunter vs Silent Spring}
\label{sec:ghunter-vs-silentspring}

We compare the effectivess of \tool{} and Silent Spring~\cite{ShcherbakovBS23} in finding universal gadgets. Silent Spring can detect prototype pollution statically and also universal gadgets in Node.js using a mix of dynamic and static taint analysis. The two approaches differ in non-trivial ways. \tool{} uses dynamic analysis to detect pollutable properties at runtime and it is driven by the test suite of a runtime environment. In contrast, Silent Spring syntactically identifies any property reads and uses them in a dynamic analysis to check if they are pollutable. This causes challenges with properties that are not identifiable statically, for example computed properties. Moreover, \tool{} analyzes all APIs systematically (subject to coverage by the test suite), while Silent Spring analyzes only 3 APIs.

Because of these differences and the fact that some of the gadgets from Silent Spring have since been fixed, we perform the following comparison: we use the gadgets identified by both toolchains as a basis for ground truth and evaluate whether or not each tool finds a gadget candidate (GC) for each \textit{property} used in the gadgets for a given API. This is because both toolchains can only taint/pollute one property at a time and report one GC per property. We focus only on ACE gadgets as was the case in Silent Spring.

\begin{table}
  \footnotesize
  \scriptsize
  \begin{tabular}{ | r | c | c | c | c | c | c | c | }
    \hline
    \multirow{2}{*}{API}      & \multirow{2}{*}{GT} & \multicolumn{3}{c}{Silent Spring} & \multicolumn{3}{|c|}{\tool{}} \\\cline{3-8}
                              &                     & GC       & TP/FP     & FN         & GC       & TP/FP   & FN       \\\hline
    \verb|cp.exec|            & 1                   & 9        & 0/9       & 1          & 2        & 1/1     & 0        \\       
    \verb|cp.execFile|        & 1                   & 9        & 0/9       & 1          & 2        & 1/1     & 0        \\       
    \verb|cp.execFileSync|    & 4                   & 11       & 3/8       & 1          & 7        & 4/3     & 0        \\       
    \verb|cp.execSync|        & 2                   & 3        & 1/2       & 1          & 3        & 2/1     & 0        \\       
    \verb|cp.fork|            & 1                   & 5        & 0/5       & 1          & 1        & 1/0     & 0        \\       
    \verb|cp.spawn|           & 3                   & 9        & 2/7       & 1          & 5        & 3/2     & 0        \\       
    \verb|cp.spawnSync|       & 4                   & 6        & 3/3       & 1          & 7        & 4/3     & 0        \\       
    \verb|import|             & 1                   & 0        & 0/0       & 1          & 1        & 1/0     & 0        \\       
    \verb|vm.SyntheticModule| & 3                   & 3        & 1/2       & 2          & 1        & 1/0     & 2        \\\hline 
    Total                     & 20                  & 55       & 10/45     & 10         & 29       & 18/11   & 2        \\\hline
  \end{tabular}
  \caption{Silent Spring vs \tool{} on Node.js v21.0.0 with properties used in \tool{} ACE gadgets as ground truth.}
  \label{tab:comparison-20}
\end{table}

Our first experiment uses the gadgets of Silent Spring as a ground truth on Node.js v16.13.1. We recreated PoCs for all its gadgets to determine the affected APIs and necessary properties. Based on this we created new test cases in the style of Silent Spring's dynamic analysis. We reran both Silent Spring and \tool{} on Node.js v16.13.1 using these new test cases to obtain the results shown in Table \ref{tab:comparison-16}. Ground truth (GT) is the number of GCs required to identify all gadgets of an API. False negatives (FN) represent the number of GCs that were identified manually (and not by a tool), but are in the GT of a gadget. We see that \tool{} is more precise (0.43 compared to 0.11) and has better recall (0.88 compared to 0.64). This is due to the underlying dynamic analysis, which guarantees that a polluted property reaches a sink. \tool{} has three FNs because it lacks features necessary to detect the sink (the \verb|require| gadget requires a chain of pollution; the \verb|vm| gadget requires array support). For Silent Spring we find nine FNs. The FNs for child process (\verb|cp|) are due to the lack of support for \verb|for-in| analysis, causing it to miss one variant of the gadgets. For \verb|import| it fails to detect the gadget API and for \verb|require| it fails to detect one property; in these cases the true and false positives would have allowed the analyst to extrapolate the properties reported as FNs here.

Our second experiment uses the gadgets of \tool{} as a ground truth on Node.js v21.0.0. For a fair comparison, we created test cases for ACE gadgets from Table \ref{tab:nodejs-gadgets} in the style of Silent Spring's dynamic analysis. We reran both \tool{} and Silent Spring on Node.js v21.0.0 using these new test cases to obtain the results shown in Table \ref{tab:comparison-20}.
For this selection of gadgets, \tool{} finds more gadgets while reporting fewer gadget candidates, again showing better precision (0.62 compared to 0.18) and recall (0.90 compared to 0.50), requiring less manual work. Silent Spring again exhibits FNs for all child process APIs because it lacks support for \verb|for-in| construct. For the \verb|import| gadget, Silent Spring fails to detect the API that triggers the gadget.

In summary, these experiments show that \tool{} is more precise, resulting in less manual work required and higher accuracy. We believe this is primarily due to the fully dynamic approach used by \tool{}, which guarantees every GC reaches a sink and provides support for dynamic language features. The shortcomings of \tool{} are due to the limitations discussed in Section \ref{sec:limitations}.

\subsection{Performance Overhead and Transparency}

We evaluated the performance overhead incurred by \tool{} in comparison with the unmodified JavaScript runtimes. To evaluate the effect of the customized runtimes and the customized V8 engines on the behavior of runtime APIs, referred to as transparency, we use the test suites as oracles to identify behavioral changes.

\tightpar{Node.js}
Running the full Node.js test suite, which contains 3,810 tests, using our modifications increased runtime by 111.72\% (from 252s to 542s). The success rate decreased from 3,782 to 3,669 cases, marking a 2.99\% reduction. The number of tests failing due to timeout increased from 2 to 44 cases.

\tightpar{Deno}
Running the three different test suites using our modifications increased runtime by 4.46\% (from 157s to 164s) for Deno core, by 43.85\% (from 130s to 187s) for Deno's Node.js compatibility module, and by 5.93\% (from 253s to 268s) for Deno std. In total that is 14.63\% (from 540s to 619s). The success rate decreased by by 0.17\% (from 1,145 to 1,143 out of 1,340) for Deno core, by nothing for Deno's Node.js compatibility module, and by 0.27\% (from 2,207 to 2,201 out of 2,258) for Deno std. In total that is 0.15\% (from 5,364 to 5,356 out of 5,648).  The number of tests failing due to timeout increased from 1 to 2 cases.

\tightpar{Evaluation} The main reason for the decreased performance and higher failure rate is the code responsible for checking tainted values in internal sinks. This code recursively traverses received values of each argument of the sink. Unexpected exceptions in the traversed objects' code, such as in property getters, lead to failures. Additionally, the modified version extends \verb|globalThis| with \verb|log|, causing some tests to fail.

\section{Defense Best Practices}
\label{sec:defense}

While previous works provide convincing evidence on the dangers of prototype pollution, as of today, there is no comprehensive defense against this vulnerability. In this section, we systematize the current proposals and mitigations and outline directions for future work.
Since our universal gadgets require the existence of prototype pollution, a reasonable question to ask is whether we should mitigate the impact of the vulnerability by fixing the gadgets. Given the lack of comprehensive defenses against prototype pollution, we think that gadgets should be treated similarly to memory corruption vulnerabilities such as return-oriented programming (ROP) and jump-oriented programming (JOP), due to their high impact.
Developers of runtimes or libraries are unaware of the presence of prototype pollution in the applications using their code. Therefore, it stands to reason to assume the presence of vulnerabilities and treat the prototype objects as untrusted data, thus guaranteeing security by fixing gadgets in their code. Similarly, application developers are unaware of prototype pollution in third-party libraries or runtimes of their application, hence they should  mitigate  gadgets. 

\subsection{Gadget Mitigations}
\label{sec:defense-guidelines}
Gadget can be mitigated by avoiding the use of potentially polluted properties in the code. 
A solution is to ensure that any access to the properties of an object does not fall back to the object's prototype chain.
We distinguish different mitigations depending on where in the code an object with a polluted prototype may be \textit{created}. This can be either the developer's own code (e.g., a library or module) or third-party code (e.g. dependencies or application code that use APIs provided by the developer). This leads us to the first guideline.

\begin{mdframed}[frametitle={G1: Explicit access to own properties}]
  If the code accesses a property in only a few instances, 
  developers should verify each access explicitly.
\end{mdframed}

Developers should check if an object defines an own property before accessing it. This can be achieved with built-in methods such as \verb|Object.hasOwn(obj, 'prop')|. We encountered this pattern regularly during our analysis of for-in loops to prevent reading unexpected properties. These checks should be added every time a potentially undefined property is accessed, thus preventing access to a polluted property.  This guideline can be applied regardless of where the object being checked was created. However, overuse of these checks increases the codebase's complexity. Therefore, developers should follow  other recommendations whenever their code makes use of many property accesses. We also recommend using the method \verb|Object.keys|, which returns the object's own enumerable properties rather than for-in loops, which additionally iterate over properties in the prototype chain. 
\
\begin{mdframed}[frametitle={G2: Safe object creation}]
  When creating an object, 
  developers should use either \verb|null| prototypes or built-in objects \verb|Map| and \verb|Set|.
\end{mdframed}

The method call \verb|Object.create(null)| and the object literal \verb|{__proto__:null}| allow to create objects that do not inherit from the prototype hierarchy. In this case, any property access \verb|obj.prop| returns \verb|undefined| unless \verb|prop| is an own property of object \verb|obj|.
On the downside, this solution can lead to unexpected exceptions. For example, code patterns like \verb|obj + "str"| will throw an exception because no \verb|toString| method is available without the prototype.

When the created object is returned by the underlying function or it is passed as an argument to a third-party function, developers should copy the object to a new object that includes \verb|Object.prototype| to ensure backward compatibility. 
We recommend assigning default values to unused properties to prevent pollution with attacker-controlled values in third-party code.
This operation can be facilitated by, e.g., using the method \verb|Object.assign({}, defaultObj, obj)|. 
We remark that the prototypes of nested objects require cloning the object by means of a deep copy algorithm, for example, using  the global method \verb|structuredClone()|.

An alternative solution is to use built-in objects that provide safe access to properties. For instance, the \verb|Map| object holds key-value pairs and provides methods such as \verb|Map.get| that do not use the prototype chain to look up the stored values. Hence, \verb|map.get('prop')| can serve as a replacement for accesses to objects. 

\begin{mdframed}[frametitle={G3: Safe copy of input data}]
  Whenever an object is received  as input data, 
  developers should copy the object's properties to a safe object.
\end{mdframed}

If a developer uses an object as a function argument  (for example, \verb|options| in Listing~\ref{lst:spawn-source}), or an object originating from a deserialization function (for example,  \verb|JSON.parse| in Listing~\ref{lst:require-source}), they should assume that the object's prototype can be polluted. A safe solution is to copy the expected properties to a new object with \verb|null| prototype. This can be achieved by creating a copy with only own properties, using the expression \verb|{__proto__:null,...obj}|.
If the code returns the received object back, the developers should use the original value instead of the copied one to avoid compatibility issues.

The guidelines G1 and G3 may be backward incompatible when an object relies on a prototype chain to define properties within nested prototypes. We expect this design pattern to be used for functions rather than data-type properties, which are subject to prototype pollution. An empirical evaluation is necessary to validate this claim.

As we can see, systematic mitigation of gadgets is an open problem.  Developers are expected to identify all gadgets to universally apply mitigation techniques to any potentially undefined property, which is infeasible in practice.
Moreover, gadget mitigation can be hard to apply to existing code bases since it requires identifying every access to undefined  properties. These considerations motivate the need for solutions like the one proposed in this paper but we believe the guidelines can be automated as suggestions for quick fixes in IDEs or similar tooling. Detection may require inter-procedural analysis, yet we expect that G1 and G2 can be implemented based on quick intra-procedural analysis.

\subsection{Prototype Pollution Mitigations}
Prototype pollution is the root cause for exploitation of gadgets, hence a comprehensive mitigation technique would solve the problem altogether. As with gadget mitigations, this requires striking a balance between security and usability, which makes it a challenging task. Here we discuss recommendations for developers and opportunities for researchers. 

\begin{table*}
  \footnotesize
  \centering
  \begin{tabular}{ |r|l|l|c|l|l|l| } 
    \hline
    \textbf{Application} & \textbf{Version} & \textbf{Vulnerability Report} & \textbf{PP Fix} & \textbf{Gadget} & \textbf{Gadget Fix} & \textbf{App Mitigations} \\\hline
    \multirow{4}{*}{Kibana}        & 6.6.0   & \href{https://research.securitum.com/prototype-pollution-rce-kibana-cve-2019-7609/}{CVE-2019-7609}      & \myCheck{} & \verb|child_process.spawn|    & \myCross{}            & \myCheckOrange{} G2, G3\myStar{} \\ \cline{2-7}
                  & 7.6.2   & \href{https://hackerone.com/reports/852613}{HackerOne \#852613} & \myCheck{} & \verb|lodash.template|        & \myCross{}            & \myCross{} \\ \cline{2-7}
                  & 7.7.0   & \href{https://hackerone.com/reports/861744}{HackerOne \#861744} & \myCheck{} & \verb|lodash.template|        & \myCross{}            & \myCheck{} G3 \\ \cline{2-7}
                  & 8.7.0   & \href{https://arxiv.org/pdf/2311.03919.pdf}{CVE-2023-31415}     & \myCheck{} & \verb|nodemailer|             & \myCross{}            & \myCross{} \\ \hline
    npm-cli       & 8.1.0   & Reported by~\cite{ShcherbakovBS23}      & \myCheck{} & \verb|child_process.spawn|    & \myCheckOrange{} G2   & \myCross{} \\ \hline
    \multirow{6}{*}{Parse Server}  & 4.10.6  & \href{https://huntr.com/bounties/ac24b343-e7da-4bc7-ab38-4f4f5cc9d099/}{CVE-2022-24760}     & \myCheck{} & \verb|bson|                   & \myCross{}            & \myCheckOrange{} Denylisting \\ \cline{2-7} 
                  & 5.3.1   & \href{https://github.com/parse-community/parse-server/security/advisories/GHSA-prm5-8g2m-24gg}{CVE-2022-39396}     & \myCheck{} & \verb|bson|                   & \myCross{}            & \myCheckOrange{} Denylisting \\ \cline{2-7}
                  & 5.3.1   & \href{https://github.com/parse-community/parse-server/security/advisories/GHSA-xprv-wvh7-qqqx}{CVE-2022-41878}     & \myCheck{} & \verb|bson|                   & \myCross{}            & \myCheckOrange{} Denylisting \\ \cline{2-7}
                  & 5.3.1   & \href{https://github.com/parse-community/parse-server/security/advisories/GHSA-93vw-8fm5-p2jf}{CVE-2022-41879}     & \myCheck{} & \verb|bson|                   & \myCross{}            & \myCheckOrange{} Denylisting \\ \cline{2-7}
                  & 5.3.1   & Reported by~\cite{ShcherbakovBS23}       & \myCheck{} & \verb|require|                & \myCheckOrange{} G2\myStar{}, G3 & \myCross{} \\ \cline{2-7} 
                  & 6.2.1   & \href{https://github.com/parse-community/parse-server/security/advisories/GHSA-462x-c3jw-7vr6}{CVE-2023-36475}     & \myCheck{} & \verb|bson|                   & \myCheck{}            & -- \\ \hline
    Rocket.Chat   & 5.1.5   & \href{https://hackerone.com/reports/1631258}{CVE-2023-23917}     & \myCheck{} & \verb|bson|                   & \myCheck{}            & -- \\ \hline
  \end{tabular}
  \caption{A summary of the RCEs exploited via prototype pollution. For each application, we list the vulnerable version, a reference to the report, and the exploited gadget. \emph{PP Fix} shows whether the prototype pollution was fixed; \emph{Gadget Fix} shows whether the gadget was fixed, including any applied guidelines; \emph{App Mitigations} details if mitigations against the attack were implemented in the application. \myCross{} indicates that no fix has been applied; \myCheckOrange{} indicates that a fix was applied but later bypassed; \myCheck{} indicates that a fix was applied and effectively protects against similar attacks. (\myStar{}) denotes a guideline that might be bypassed.} 
  \label{tab:eval-apps}
\end{table*}

\tightpar{Guidelines for developers}
A general solution is to prevent any accesses to the prototypes of objects, which can be achieved by the above-mentioned guidelines for gadget mitigation.  Following  guideline G1, developers should avoid accesses to object prototypes through property reading expressions. This is because properties such as \verb|__proto__| and \verb|constructor.prototype|, which give accesses to the prototype chain, are not defined in the object itself.  Alternatively, this can also be achieved by explicitly checking accesses to properties  \verb|__proto__|, \verb|constructor|, and \verb|prototype|. Similar to own property checks for gadget mitigation, this mitigation introduces additional verbosity.
Following guideline G2, one can instead use data structures with either \verb|null| prototypes or safe \verb|get| and \verb|set| functions. 

Another solution is to prevent unintended modification to the prototype object itself, which can be achieved with built-in functions such as \verb|freeze|, \verb|preventExtension|, and \verb|seal|~\cite{object-mdn}. These functions offer a mechanism to prevent the creation of new properties on an object. The \verb|freeze| function additionally prevents overwriting. Node.js provides the experimental command-line feature, \verb|--frozen-intrinsics|, which freezes the prototypes of built-in objects like \verb|Array| and \verb|Object|. Similarly, Deno removes \verb|__proto__| from \verb|Object.prototype| by default.

While mitigating prototype pollution, these solutions can be problematic for third-party packages that rely on changing the prototype to implement, e.g., polyfills. Also, they require coverage of all prototype object, including user-defined classes which makes it verbose and hard to maintain for large projects. We recommend these solution for the development of a new project while existing project should perform regression testing to ensure that no functionalities are disrupted.

\tightpar{Research opportunities}
Mitigation of prototype pollution and gadgets remains an open problem. A recent proposal driven by Google aims to prevent prototype pollution at the language- and runtime-level \cite{proposal}. It proposes an opt-in \textit{secure mode}, which, if enabled, prevents accesses to prototypes with dynamic string keys. It allows prototype access through reflection APIs instead of strings, thus only requiring changes to \verb|__proto__| and \verb|constructor|, whenever they are accessed purposefully. While an important step in the right direction, this solution poses challenges of backward compatibility for server- and client-side applications.

\subsection{Case Studies}
\label{sec:case-studies}

We evaluate fixes of known server-side prototype pollution vulnerabilities and their gadgets to identify common issues in mitigations that permit attackers to bypass the fixes. We conducted our search through public vulnerability reports on HackerOne, blog posts, and publications related to open-source applications over the past 5 years, summarizing our findings in Table~\ref{tab:eval-apps}. Our results contain 12 exploitable cases leading to Remote Code Execution (RCE) in 4 popular applications. The root cause of their exploitability, namely code patterns that allow to pollute prototypes, has been addressed in all cases. These vulnerabilities involve 5 unique gadgets to achieve RCEs. For 4 of these gadgets, developers proposed either fixes or mitigations for the attacks.

We identify 6 vulnerabilities that exploit a gadget in the \verb|bson| package.
The Parse Server developers fixed 5 vulnerabilities that use this gadget  with input data validation through denylisting. However, these mitigations were bypassed several times through unexpected means, e.g. with files metadata. Ultimately, the dangerous feature was removed from \verb|bson|, thereby fixing the gadget. Both Parse Server and Rocket.Chat  fixed their vulnerabilities through this method. This highlights the need to fix gadgets because mitigation is difficult and often leaves room for exploitation by other means.

The gadgets in \verb|lodash.template| and \verb|nodemailer| remain unaddressed and could be exploited given new prototype pollutions. The maintainers of Kibana banned the use of \verb|lodash.template| in their code and mitigated it by intercepting \verb|template| calls and validating the  polluted property when the package is included as a transitive dependency.

However, as illustrated, it can be dangerous to leave gadgets unfixed. Next, we detail two interesting gadgets and highlight issues in their fixes to demonstrate the risk.

\tightpar{child\_process.spawn}
The first mention of the \verb|spawn| gadget appears in the report CVE-2019-7609 by Michał Bentkowski, outlining a prototype pollution vulnerability in Kibana. Kibana spawns a \verb|node| process, and the security researcher discovered a method to execute arbitrary code through crafted environment variables of the new process.

\begin{figure}[t]
\centering
\begin{lstlisting}[caption={Simplified Node.js \textit{spawn} implementation.},label={lst:spawn-source}]
function spawn(file, args, options) {
  if (options === undefined)
    options = {}
  options = Object.assign({}, options)
  options.env = options.env || process.env
  options.file = options.shell || file
  //...
  internalSpawn({
    file: options.file,
    env: options.env,
    //...
  })
}
\end{lstlisting}
\end{figure}

Listing~\ref{lst:spawn-source} presents the necessary code of the \verb|spawn| function to understand the attack. If an application invokes \verb|spawn| with two arguments, \verb|file| and \verb|args|, then the third argument \verb|options| is undefined.
Line 3 creates a new object that inherits \verb|Object.prototype|, making it susceptible to prototype pollution.    
Line 4 makes a shallow copy of \verb|options| to prevent changing the user's options object if passed. In our scenarios, this copy operation is inconsequential because \verb|options| is an empty object created within the function itself.
Line 5 retrieves the value of the \verb|env| property. If the value is undefined, the code defaults to \verb|process.env|, assigning this to the \verb|env| property of options. 
Line 6 similarly handles the \verb|shell| property from options and the \verb|file| parameter.
Subsequently, the code passes the aggregated options to the internal implementation of the \verb|spawn| function, which initiates a new process. If an attacker pollutes the \verb|env| property in \verb|Object.prototype|, line 5 will read the attacker-controlled value instead of system environment variables. It allows the attacker to execute arbitrary code, leading to RCE in Kibana.

The Kibana team fixed the prototype pollution vulnerability and mitigated the gadget in \href{https://github.com/elastic/kibana/pull/55697}{PR \#55697} to prevent similar attacks in later versions.
Because the gadget is part of Node.js' source code, application developers are limited to intercepting \verb|spawn| calls and altering the arguments. Listing~\ref{lst:spawn-mitigation} provides a simplified version of this mitigation. The code uses a JavaScript Proxy to invoke the \verb|patch| function, thereby securing the options. It evaluates passed arguments from the zero-based array \verb|args|. If the argument at position 1 is an array, line 5 simply advances the position. If the subsequent argument at position 2 is an object, it is treated as the options, and the \verb|prototypeless| function then copies the options' own properties to new objects with null prototypes. 

\begin{figure}[t]
\centering
\begin{lstlisting}[caption={Simplified \textit{spawn} gadget mitigation in Kibana.},label={lst:spawn-mitigation}]
cp.spawn = new Proxy(cp.spawn, {apply: patch})
function patch(target, thisArg, args) {
  var pos = 1;
  if (Array.isArray(args[pos]))
    pos++ // fn(file, args, ...)
  if (typeof args[pos] === 'object') {
    // fn(file, options, ...)
    // fn(file, args, options, ...)
    args[pos] = prototypeless(args[pos])
  }
  //...
  return target.apply(thisArg, args)
}
function prototypeless(obj) {
  var newObj = Object.assign(
    Object.create(null), obj)
  newObj.env = Object.assign(
    Object.create(null), newObj.env)
  return newObj
}
\end{lstlisting}
\end{figure}

This mitigation follows  our guidelines G2 and G3. Lines 16 and 18 create new objects with null prototypes in accordance with G2, ensuring that care is also taken for nested objects to prevent pollution of \verb|env| when the value is read from \verb|process.env|.
The use of \verb|Object.assign| in lines 15 and 17 copies only own properties from the original objects to the new objects with null prototypes, following  G3.

However, this mitigation has two critical weaknesses that allow the attacker to bypass it. Developers are constrained to validating arguments and lack control over modifications to arguments after passing them to Node.js functions. As observed in line 5 of Listing~\ref{lst:spawn-source}, the \verb|spawn| function makes a copy of the received options into a common empty object that shares its prototype with others. Consequently, any properties of the options might be polluted again. Fortunately, \verb|spawn| does not copy the \verb|env| property, so environment variables are not affected. The other weakness is more dangerous and allows for bypassing all mitigations and even security fixes in Node.js, as we will see later. Lines 6 and 9 of Listing~\ref{lst:spawn-mitigation} are also exploitable by prototype pollution. The array \verb|args|, like any array, has \verb|Object.prototype| in its prototype chain and looks up an undefined property. Therefore, polluting the property \verb|2| allows the attacker to control the options. For this exploit, a gadget trigger might look as follows:

\begin{lstlisting}
Object.prototype[2] = { env: 
  {NODE_OPTIONS: '--inspect-brk=0.0.0.0:1337'}
}
spawn('node', ['any_file.js'])
\end{lstlisting}

Thus, the \verb|spawn| gadget is still exploitable in Kibana after mitigations. This case highlights the importance for developers to exercise caution with security-critical code, such as gadget mitigations, and to test it against other gadgets using tools like \tool{} to avoid introducing new exploitation flows into the code.

Shcherbakov et al.~\cite{ShcherbakovBS23} introduce a variation of the \verb|spawn| gadget. They find that the name of a running process can be manipulated through the polluted property \verb|shell|, as shown in line 6 of Listing~\ref{lst:spawn-source}. Additionally, they disclose new payloads for the exploit that operate without controlling environment variables and controlling only one variable. They identify a vulnerability in  the JavaScript package manager npm-cli, and exploit it to demonstrate the practical feasibility of using this gadget. Although npm-cli contributors addressed the reported prototype pollution, they did not mitigate the gadget.

In June 2022, the Node.js team attempted to fix this gadget in \href{https://github.com/nodejs/node/pull/43159}{PR \#43159}.
In terms of our terminology, they implemented guideline G2 by assigning the value \verb|ObjectFreeze(ObjectCreate(null))| to options in line 3 of Listing~\ref{lst:spawn-source} and eliminated \verb|Object.assign()| in line 4 to maintain the usage of options with a null prototype. As discussed in Section~\ref{sec:defense-guidelines}, G2 alone is insufficient to prevent all forms of gadget exploitation, and G2 should be used in conjunction with G3. \tool{} reports a gadget  for \verb|spawn| when a user supplies their own options object to \verb|spawn|:

\begin{lstlisting}
Object.prototype.shell = 'node'
Object.prototype.env = 
  {NODE_OPTIONS: '--inspect-brk=0.0.0.0:1337'}
spawn('app', ['file.log'], {cwd: '/tmp'})
\end{lstlisting}

This case illustrates the importance of a consistent approach in implementing gadget fixes. When applying  guideline G2, it is crucial to carefully handle input data and copy it safely, while also applying G3. Relying on validating security-critical parameters outside the gadget proves to be insecure. 

\begin{figure}[t]
\centering
\begin{lstlisting}[caption={Simplified Node.js \textit{require} implementation.},label={lst:require-source}]
// lib\internal\modules\cjs\loader.js
function readPackage(dir) {
  const jsonPath = resolve(dir, 'package.json')
  const json = packageJsonReader.read(jsonPath)
  if (json === undefined)
    return false
  return JSON.parse(json)
}
function tryPackage(requestPath) {
  const pkg = readPackage(requestPath)?.main
  if (!pkg) {
    const js = resolve(requestPath, 'index.js')
    return loadFile(js)
  }
  loadFile(pkg)
}
\end{lstlisting}
\end{figure}

%
%

\tightpar{require}
Shcherbakov et al.~\cite{ShcherbakovBS23} report a gadget in \verb|require|, a built-in function in Node.js for including external modules from separate files as well as Node.js modules, and utilize this gadget in one of the Parse Server exploits.
Listing~\ref{lst:require-source} illustrates a gadget based on simplified Node.js code. The function \verb|tryPackage| receives a directory path for a module and invokes \verb|readPackage()| in line 10. The code in line 4 attempts to read \verb|package.json| from the given directory. If the read operation is successful, \verb|readPackage()| parses the content of the file as JSON and returns the parsed object in line 7. \verb|tryPackage| then accesses the \verb|main| property in line 10, loads a file based on the path specified in the \verb|main| property, and evaluates its JavaScript code in line 15. 
Consequently, if \verb|package.json| lacks the main property, line 10 looks up the property in the prototype chain of the returned object, allowing a polluted property from \verb|Object.prototype| to be assigned to \verb|pkg|. 
This leads to the evaluation of JavaScript code from an attacker-controlled file in line 15.

The Node.js team attempted to fix this gadget by applying guidelines G2 and G3 to \verb|readPackage| function. They correctly make a safe copy of the parsed object in line 7 to an object with a null prototype. 
However, \tool{} detects a variation of the gadget in v18.13.0. If \verb|packageJsonReader| can not find the \verb|package.json| file, the function returns \verb|false| in line 6. Since Boolean is a primitive type and all primitive types in JavaScript inherit from \verb|Object.prototype|, the expression \verb|(false)?.main| in line 10 accesses the polluted value in \verb|Object.prototype| and assigns it to \verb|pkg|, achieving the same attack. This makes the \verb|require| function exploitable, albeit through a different gadget.

\tightpar{End-to-end exploit} 
To demonstrate the impact of this gadget, we analyze Kibana version 8.7.0 for end-to-end exploits. We initially utilized the Silent Spring~\cite{ShcherbakovBS23} toolchain to detect prototype pollution vulnerabilities. The analysis reports 44 cases in the server-side code, with 6 being potentially exploitable. 
The simplified code of one of the cases is presented in Listing~\ref{lst:kibana-pp}. Kibana loads a config file, parses it into an object, and expands the properties from dot notation into nested objects (e.g., \verb|{a.b:0}| to \verb|{a:{b:0}}|) with the \verb|ensureDeepObject| function. This code is vulnerable to prototype pollution. On line 19, the first argument allows an attacker to get a reference to the prototype and then assign a value to any property of the prototype in line 14.

\begin{figure}[t]
\centering
\begin{lstlisting}[caption={Prototype pollution vulnerability in Kibana.},label={lst:kibana-pp}]
function ensureDeepObject(obj: any): any {
  return Object.keys(obj).reduce((res, key)=>{
    const val = obj[key];
    if (!key.includes('.'))
      res[key] = ensureDeepObject(val);
    else
      walk(res, key.split('.'), val);
    return res;
  }, {} as any);
}
function walk(obj:any, keys:string[], val:any){
  const key = keys.shift()!;
  if (keys.length === 0) {
    obj[key] = val;
    return;
  }
  if (obj[key] === undefined)
    obj[key] = {};
  walk(obj[key], keys, ensureDeepObject(val));
}
\end{lstlisting}
\end{figure}

To exploit this prototype pollution, an attacker should upload a configuration file with a payload via the Web UI form and restart Kibana to trigger the parsing of the new configuration file.
During the restart process, Kibana crashed at an early stage due to an unexpected polluted property that prevented gadget execution via another web request. However, the application invoked \verb|require| multiple times during loading, allowing us to trigger it and achieve RCE. The investigation process took 8 hours for one author already familiar with Kibana. We reported this vulnerability, and the Kibana team acknowledged the issue, assigning CVE-2023-31414 with a critical CVSS  9.1, and rewarding a substantial bounty. The Node.js team fixed the \verb|require| gadget in version 18.19.0.

\tightpar{Takeaways}
If developers fix only the prototype pollution vulnerabilities while leaving its associated gadget exploitable, they remain at risk. Our case studies show that many developers are aware of this risk and attempt to mitigate the gadgets and similar attacks. However, this task is far from trivial. We identified numerous gadgets and common coding issues that lead to new gadgets, emphasizing the need for more principled solutions. Our proposed guidelines are a step forward in this direction.

\section{Related Work}

We discuss our work in the context of closely-related works that address prototype pollution vulnerabilities and position our contributions in the area
of web application security.

\tightpar{Universal gadgets in JavaScript runtimes} 
The problem of identifying universal gadgets in JavaScript runtimes remains largely unexplored. To the best of our knowledge, only the work of Shcherbakov et al.  \cite{ShcherbakovBS23} studies universal gadgets in Node.js. Section \ref{sec:ghunter-vs-silentspring} compares their work to \tool.

Recent  work by Shcherbakov et al. \cite{shcherbakov2024unveiling} uses dynamic taint analysis via program instrumentation to find gadgets in NPM packages. This approach cannot be used to identify universal gadgets which require modifications of runtime environments (Node.js and Deno) and  the underlying V8 engine. Our universal gadgets are complementary and contribute with additional dangerous sinks for analysis such as \cite{shcherbakov2024unveiling}, thus increasing their attack surface coverage. Kang et al.~\cite{Kang22}  study prototype pollution on the client-side application  by dynamic taint tracking. Their analysis is implemented at the V8 JavaScript engine by adapting the tool of  Melicher et al. \cite{domxss:ndss18}.
Their focus on client-side vulnerabilities is incompatible with server-side runtimes such as Node.js and Deno. 

Other work \cite{liu2024undefined,steffens2021understanding} uses concolic execution to find gadgets in client-side JavaScript code. Concolic execution is a promising enhancement of dynamic analysis. Liu et al. \cite{liu2024undefined} focus specifically on finding gadget chains where one gadget unlocks the use of another gadget (e.g. by forcing a branch). It would be interesting to apply these ideas to backend systems.

\tightpar{Prototype pollution} In recent years, we have seen increased attention on prototype pollution vulnerabilities by both academia and  practitioners \cite{arteau2018prototype,kim2021dapp,Li21,Li22,Kang22,ShcherbakovBS23,blackbox,pp-finder,blackfan}.  The work of Arteau \cite{arteau2018prototype} is the first to demonstrate the feasibility of prototype pollution in a number of libraries.  On the academic front,  the vast majority of research contributions  focus on the detection of prototype pollution~\cite{Li21,Li22,kim2021dapp}. These works use static taint analysis to find zero-day vulnerabilities leading to DoS attacks.  Our contributions are complementary as they focus on the detection of universal gadgets rather than prototype pollution.  The security  impact of prototype pollution is discussed in practitioner forums~\cite{blackbox,pp-finder,blackfan}. Heyes \cite{blackbox}  describes how prototype pollution can be exploited in Node.js to find vulnerabilities beyond DoS in black-box scenarios. Their semi-automated approach uses  PP-finder \cite{pp-finder} to report all undefined properties encountered during the execution and conducts manual inspection of packages for vulnerabilities. This approach is practical for a few specific targets, yet it is neither feasible at scale nor able to identify universal gadgets. 

\tightpar{Code reuse attacks for the web}
Prototype pollution is a new class of  code reuse vulnerabilities in web applications and, as such, it shares similarities with object injection vulnerabilities. Several works use static taint analysis to detect code reuse vulnerabilities for a variety of languages including 
PHP~\cite{esser2010utilizing,Dahse14,Dahse14Usenix,ParkKJS22}, .NET~\cite{bh17,ShcherbakovB21},  and  Java~\cite{munoz2018serial,10.1145/2976749.2978361}. Xiao et al.~\cite{Xiao21} study a related type of vulnerability coined  hidden property attacks. Lekies et al.~\cite{LekiesKGNJ17} and Roth et al.~\cite{Roth0S20}  study the implications of script gadgets in  bypassing existing XSS and CSP mitigations. While all of these vulnerabilities rely on the reuse of code gadgets, their precise connection is yet to be studied systematically. 
\tool{}  implements a lightweight form of dynamic taint analysis at the level of JavaScript runtimes and V8 engine. Dynamic taint analysis~\cite{DBLP:conf/sp/SchwartzAB10,explicitsecrecy}  is a popular technique  used to identify web-related vulnerabilities, including  instrumentations at both program- and runtime-level \cite{jalangi,Lekies:2013:MFL:2508859.2516703,ichnea,GauthierHJ18,NielsenHG19,vv8-imc19,CasselWJ23,abbadini2023cage4deno}.   

\section{Conclusion}

We have presented a semi-automated pipeline, \tool{}, able to find exploitable universal gadgets in Node.js and Deno by lightweight dynamic taint analysis. We have used \tool{} in a comprehensive study of universal gadgets, finding a total 123 exploitable gadgets. In absence of comprehensive defenses, we have systematized existing mitigation for prototype pollution and gadgets in the form of guidelines. We have used these guidelines in a study of existing exploits in real applications to illuminate the current status, finding a high-severity exploit due to the lack of principled mitigations. 

\paragraph{Acknowledgments}
We thank anonymous reviewers for the helpful suggestions and feedback.  
This work was partially supported by the Swedish Foundation for Strategic
Research (SSF) under project CHAINS, the Swedish Research Council (VR) under project WebInspector, and 
Wallenberg AI, Autonomous Systems and Software Program (WASP)
funded by the Knut and Alice Wallenberg Foundation under project ShiftLeft. 




\bibliographystyle{plain}
\bibliography{references}

\begin{thebibliography}{10}

\bibitem{nodejs-fast-api}
Adding v8 fast api.
\newblock
  \url{https://github.com/nodejs/node/blob/v21.0.0/doc/contributing/adding-v8-fast-api.md}.

\bibitem{blackfan}
{Client-Side Prototype Pollution and useful Script Gadgets}.
\newblock \url{https://github.com/BlackFan/client-side-prototype-pollution}.

\bibitem{Deno}
{Deno, the next-generation JavaScript runtime}.
\newblock \url{https://deno.com/}.

\bibitem{Nodejs}
{Node.js JavaScript runtime}.
\newblock \url{https://nodejs.org/}.

\bibitem{object-mdn}
{Object - JavaScript - MDN}.
\newblock
  \url{https://developer.mozilla.org/en-US/docs/Web/JavaScript/Reference/Global_Objects/Object}.

\bibitem{proposal}
Prototype pollution mitigation / symbol.proto.
\newblock \url{https://github.com/tc39/proposal-symbol-proto}.

\bibitem{ECMA335}
Standard ecma-335 common language infrastructure (cli).
\newblock
  \url{https://www.ecma-international.org/publications/standards/Ecma-335.htm}.

\bibitem{abbadini2023cage4deno}
Marco Abbadini, Dario Facchinetti, Gianluca Oldani, Matthew Rossi, and Stefano
  Paraboschi.
\newblock Cage4deno: A fine-grained sandbox for deno subprocesses.
\newblock 2023.

\bibitem{AhmadpanahHBOS21}
Mohammad~M. Ahmadpanah, Daniel Hedin, Musard Balliu, Lars~Eric Olsson, and
  Andrei Sabelfeld.
\newblock {SandTrap}: Securing {JavaScript}-driven trigger-action platforms.
\newblock In {\em {USENIX} Security Symposium}, 2021.

\bibitem{arteau2018prototype}
Olivier Arteau.
\newblock Prototype pollution attack in {NodeJS} application.
\newblock {\em NorthSec}, 2018.

\bibitem{BrownNWEJS17}
Fraser Brown, Shravan Narayan, Riad~S. Wahby, Dawson~R. Engler, Ranjit Jhala,
  and Deian Stefan.
\newblock Finding and preventing bugs in {JavaScript} bindings.
\newblock In {\em Symposium on Security and Privacy ({S\&P})}, 2017.

\bibitem{v8-inline-caches}
Mathias Bynens.
\newblock Javascript engine fundamentals: Shapes and inline caches.
\newblock \url{https://mathiasbynens.be/notes/shapes-ics}.

\bibitem{CasselWJ23}
Darion Cassel, Wai~Tuck Wong, and Limin Jia.
\newblock Nodemedic: End-to-end analysis of node.js vulnerabilities with
  provenance graphs.
\newblock In {\em 8th {IEEE} European Symposium on Security and Privacy,
  EuroS{\&}P 2023, Delft, Netherlands, July 3-7, 2023}. {IEEE}, 2023.

\bibitem{artifact}
Eric Cornelissen, Mikhail Shcherbakov, and Musard Balliu.
\newblock Ghunter: Universal prototype pollution gadgets in javascript
  runtimes.
\newblock \url{https://github.com/KTH-LangSec/ghunter}.

\bibitem{Dahse14Usenix}
Johannes Dahse and Thorsten Holz.
\newblock Static detection of second-order vulnerabilities in web applications.
\newblock In {\em {USENIX} Security 14}, 2014.

\bibitem{Dahse14}
Johannes Dahse, Nikolai Krein, and Thorsten Holz.
\newblock Code reuse attacks in {PHP:} automated {POP} chain generation.
\newblock In {\em Conference on Computer and Communications Security ({CCS})},
  2014.

\bibitem{duantowards}
Ruian Duan, Omar Alrawi, Ranjita~Pai Kasturi, Ryan Elder, Brendan
  Saltaformaggio, and Wenke Lee.
\newblock Towards measuring supply chain attacks on package managers for
  interpreted languages.
\newblock In {\em Network and Distributed System Security Symposium ({NDSS})},
  2021.

\bibitem{esser2010utilizing}
Stefan Esser.
\newblock {Utilizing Code Reuse/ROP in PHP Application Exploits}.
\newblock {\em Proceedings of the Black Hat USA}, 2010.

\bibitem{GauthierHJ18}
Fran{\c{c}}ois Gauthier, Behnaz Hassanshahi, and Alexander Jordan.
\newblock {AFFOGATO:} runtime detection of injection attacks for node.js.
\newblock In {\em {International Symposium on Software Testing and Analysis
  ({ISSTA})}}, 2018.

\bibitem{gadgetsKTH}
Language-Based~Security group at KTH Royal Institute~of Technology.
\newblock Server-side prototype pollution gadgets.
\newblock \url{https://github.com/KTH-LangSec/server-side-prototype-pollution},
  2024.

\bibitem{blackbox}
Gareth Heyes.
\newblock Server-side prototype pollution: Black-box detection without the dos.
\newblock
  \url{https://portswigger.net/research/server-side-prototype-pollution}.

\bibitem{10.1145/2976749.2978361}
Philipp Holzinger, Stefan Triller, Alexandre Bartel, and Eric Bodden.
\newblock An in-depth study of more than ten years of java exploitation.
\newblock In {\em Conference on Computer and Communications Security ({CCS})},
  2016.

\bibitem{vv8-imc19}
Jordan Jueckstock and Alexandros Kapravelos.
\newblock {VisibleV8: In-browser Monitoring of JavaScript in the Wild}.
\newblock In {\em {Proceedings of the ACM Internet Measurement Conference
  (IMC)}}, October 2019.

\bibitem{Kang22}
Zifeng Kang, Song Li, and Yinzhi Cao.
\newblock Probe the proto: Measuring client-side prototype pollution
  vulnerabilities of one million real-world websites.
\newblock In {\em Network and Distributed System Security Symposium ({NDSS}
  2022)}, 2022.

\bibitem{ichnea}
Rezwana Karim, Frank Tip, Alena Sochůrková, and Koushik Sen.
\newblock Platform-independent dynamic taint analysis for javascript.
\newblock {\em IEEE Transactions on Software Engineering}, 46(12), 2020.

\bibitem{kim2021dapp}
Hee~Yeon Kim, Ji~Hoon Kim, Ho~Kyun Oh, Beom~Jin Lee, Si~Woo Mun, Jeong~Hoon
  Shin, and Kyounggon Kim.
\newblock Dapp: automatic detection and analysis of prototype pollution
  vulnerability in {Node.js} modules.
\newblock {\em International Journal of Information Security}, 2021.

\bibitem{LekiesKGNJ17}
Sebastian Lekies, Krzysztof Kotowicz, Samuel Gro{\ss}, Eduardo A.~Vela Nava,
  and Martin Johns.
\newblock Code-reuse attacks for the web: Breaking cross-site scripting
  mitigations via script gadgets.
\newblock In {\em Conference on Computer and Communications Security ({CCS})},
  2017.

\bibitem{Lekies:2013:MFL:2508859.2516703}
Sebastian Lekies, Ben Stock, and Martin Johns.
\newblock 25 million flows later: large-scale detection of {DOM}-based {XSS}.
\newblock In {\em Conference on Computer and Communications Security ({CCS})},
  2013.

\bibitem{Li21}
Song Li, Mingqing Kang, Jianwei Hou, and Yinzhi Cao.
\newblock Detecting {Node.js} prototype pollution vulnerabilities via object
  lookup analysis.
\newblock In {\em Proceedings of the 29th ACM Joint Meeting on European
  Software Engineering Conference and Symposium on the Foundations of Software
  Engineering}, ESEC/FSE 2021, 2021.

\bibitem{Li22}
Song Li, Mingqing Kang, Jianwei Hou, and Yinzhi Cao.
\newblock Mining {Node.js} vulnerabilities via object dependence graph and
  query.
\newblock In {\em {USENIX} Security Symposium}, 2022.

\bibitem{liu2024undefined}
Zhengyu Liu, Kecheng An, and Yinzhi Cao.
\newblock Undefined-oriented programming: Detecting and chaining prototype
  pollution gadgets in node. js template engines for malicious consequences.
\newblock In {\em 2024 IEEE Symposium on Security and Privacy (SP)}. IEEE
  Computer Society, 2024.

\bibitem{domxss:ndss18}
William Melicher, Anupam Das, Mahmood Sharif, Lujo Bauer, and Limin Jia.
\newblock Riding out {DOMsday}: {T}oward detecting and preventing {DOM}
  cross-site scripting.
\newblock In {\em {NDSS} 2018}, 2018.

\bibitem{bh17}
Alvaro Mu{\~n}oz and Oleksandr Mirosh.
\newblock {Friday the 13th json attacks}.
\newblock {\em Proceedings of the Black Hat USA}, 2017.

\bibitem{munoz2018serial}
Alvaro Mu{\~n}oz and Christian Schneider.
\newblock Serial killer: Silently pwning your java endpoints, 2018.

\bibitem{NielsenHG19}
Benjamin~Barslev Nielsen, Behnaz Hassanshahi, and Fran{\c{c}}ois Gauthier.
\newblock Nodest: feedback-driven static analysis of node.js applications.
\newblock In {\em Joint Meeting on European Software Engineering Conference and
  Symposium on the Foundations of Software Engineering, ({FSE})}, 2019.

\bibitem{sarif}
OASIS.
\newblock Static analysis results interchange format (sarif) version 2.1.0.
\newblock
  \url{https://docs.oasis-open.org/sarif/sarif/v2.1.0/sarif-v2.1.0.html}.

\bibitem{ParkKJS22}
Sunnyeo Park, Daejun Kim, Suman Jana, and Sooel Son.
\newblock {FUGIO:} automatic exploit generation for {PHP} object injection
  vulnerabilities.
\newblock In {\em 31st {USENIX} Security Symposium, {USENIX} Security 2022,
  Boston, MA, USA, August 10-12, 2022}.

\bibitem{Roth0S20}
Sebastian Roth, Michael Backes, and Ben Stock.
\newblock Assessing the impact of script gadgets on {CSP} at scale.
\newblock In {\em Asia Conference on Computer and Communications Security,
  ({ASIA} {CCS})}, 2020.

\bibitem{explicitsecrecy}
D.~Schoepe, M.~Balliu, B.~C. Pierce, and A.~Sabelfeld.
\newblock Explicit secrecy: A policy for taint tracking.
\newblock In {\em EuroS\&P}, 2016.

\bibitem{DBLP:conf/sp/SchwartzAB10}
Edward~J. Schwartz, Thanassis Avgerinos, and David Brumley.
\newblock All you ever wanted to know about dynamic taint analysis and forward
  symbolic execution (but might have been afraid to ask).
\newblock In {\em {IEEE} S\&P}, 2010.

\bibitem{jalangi}
Koushik Sen, Swaroop Kalasapur, Tasneem Brutch, and Simon Gibbs.
\newblock Jalangi: A selective record-replay and dynamic analysis framework for
  javascript.
\newblock In {\em Proceedings of the 37th IEEE/ACM International Conference on
  Automated Software Engineering}, ASE '22, New York, NY, USA, 2013.

\bibitem{ShcherbakovB21}
Mikhail Shcherbakov and Musard Balliu.
\newblock {SerialDetector: Principled and Practical Exploration of Object
  Injection Vulnerabilities for the Web}.
\newblock In {\em 28th Annual Network and Distributed System Security
  Symposium, {NDSS} 2021, virtually, February 21-25, 2021}, 2021.

\bibitem{ShcherbakovBS23}
Mikhail Shcherbakov, Musard Balliu, and Cristian{-}Alexandru Staicu.
\newblock Silent spring: Prototype pollution leads to remote code execution in
  node.js.
\newblock In {\em 32nd {USENIX} Security Symposium, {USENIX} Security 2023,
  Anaheim, CA, USA, August 9-11, 2023}. {USENIX} Association, 2023.

\bibitem{shcherbakov2024unveiling}
Mikhail Shcherbakov, Paul Moosbrugger, and Musard Balliu.
\newblock Unveiling the invisible: Detection and evaluation of prototype
  pollution gadgets with dynamic taint analysis.
\newblock In {\em Proceedings of the ACM on Web Conference 2024}, WWW '24, New
  York, NY, USA, 2024. Association for Computing Machinery.

\bibitem{StaicuPL18}
Cristian{-}Alexandru Staicu, Michael Pradel, and Benjamin Livshits.
\newblock {SYNODE:} understanding and automatically preventing injection
  attacks on {Node.js}.
\newblock In {\em Network and Distributed System Security Symposium ({NDSS})},
  2018.

\bibitem{StaicuSBPS19}
Cristian{-}Alexandru Staicu, Daniel Schoepe, Musard Balliu, Michael Pradel, and
  Andrei Sabelfeld.
\newblock An empirical study of information flows in real-world {JavaScript}.
\newblock In {\em 14th {ACM} {SIGSAC} Workshop on Programming Languages and
  Analysis for Security, {{PLAS}}}, 2019.

\bibitem{steffens2021understanding}
Marius Steffens.
\newblock Understanding emerging client-side web vulnerabilities using dynamic
  program analysis.
\newblock 2021.

\bibitem{StockJS017}
Ben Stock, Martin Johns, Marius Steffens, and Michael Backes.
\newblock How the web tangled itself: Uncovering the history of client-side web
  (in)security.
\newblock In {\em 26th {USENIX} Security Symposium, {USENIX} Security 2017,
  Vancouver, BC, Canada, August 16-18, 2017}. {USENIX} Association, 2017.

\bibitem{Xiao21}
Feng Xiao, Jianwei Huang, Yichang Xiong, Guangliang Yang, Hong Hu, Guofei Gu,
  and Wenke Lee.
\newblock Abusing hidden properties to attack the {Node.js} ecosystem.
\newblock In {\em {USENIX} Security Symposium}, 2021.

\bibitem{pp-finder}
YesWeHack.
\newblock Server side prototype pollution, how to detect and exploit.
\newblock
  \url{https://blog.yeswehack.com/talent-development/server-side-prototype-pollution-how-to-detect-and-exploit/}.

\bibitem{ZimmermannSTP19}
Markus Zimmermann, Cristian{-}Alexandru, Cam Tenny, and Michael Pradel.
\newblock Small world with high risks: {A} study of security threats in the npm
  ecosystem.
\newblock In {\em {USENIX} Security Symposium}, 2019.

\end{thebibliography}


\appendix
\section{Appendix}

\begin{table}
  \footnotesize
  \centering
  \begin{tabular}{|r|l|l|}
    \hline
    \textbf{Gadget}                                  & \textbf{Properties}                                                                                                                                       & \textbf{Attack Type}          \\\hline

    cluster.fork                                     & \verb|NODE_OPTIONS|                                                                                                                                       & ACE                           \\\hline
    cp.exec                                          & \verb|NODE_OPTIONS|                                                                                                                                       & ACE                           \\\hline
    cp.execFile                                      & \verb|NODE_OPTIONS|                                                                                                                                       & ACE                           \\\hline
    \multirow{5}{*}{cp.execFileSync}                 & \verb|shell|, \verb|NODE_OPTIONS|                                                                                                                         & ACE                           \\\cline{2-3}
                                                     & \verb|shell|, \verb|input|                                                                                                                                & ACE                           \\\cline{2-3}
                                                     & \verb|uid|                                                                                                                                                & PE                            \\\cline{2-3}
                                                     & \verb|gid|                                                                                                                                                & PE                            \\\cline{2-3}
                                                     & \verb|cwd|                                                                                                                                                & PT                            \\\hline
    \multirow{2}{*}{cp.execSync}                     & \verb|NODE_OPTIONS|                                                                                                                                       & ACE                           \\\cline{2-3}
                                                     & \verb|input|                                                                                                                                              & ACE                           \\\hline
    cp.fork                                          & \verb|NODE_OPTIONS|                                                                                                                                       & ACE                           \\\hline
    \multirow{4}{*}{cp.spawn}                        & \verb|shell|, \verb|NODE_OPTIONS|                                                                                                                         & ACE                           \\\cline{2-3}
                                                     & \verb|uid|                                                                                                                                                & PE                            \\\cline{2-3}
                                                     & \verb|gid|                                                                                                                                                & PE                            \\\cline{2-3}
                                                     & \verb|cwd|                                                                                                                                                & PT                            \\\hline
    \multirow{5}{*}{cp.spawnSync}                    & \verb|shell|, \verb|NODE_OPTIONS|                                                                                                                         & ACE                           \\\cline{2-3}
                                                     & \verb|shell|, \verb|input|                                                                                                                                & ACE                           \\\cline{2-3}
                                                     & \verb|uid|                                                                                                                                                & PE                            \\\cline{2-3}
                                                     & \verb|gid|                                                                                                                                                & PE                            \\\cline{2-3}
                                                     & \verb|cwd|                                                                                                                                                & PT                            \\\hline
    crypto.privateEncrypt                            & \verb|padding|                                                                                                                                            & CD                            \\\hline
    crypto.publicEncrypt                             & \verb|padding|                                                                                                                                            & CD                            \\\hline
    crypto.subtle.encrypt                            & \verb|kty|                                                                                                                                                & Segfault                      \\\hline
    crypto.publicKey.export                          & \verb|kty|                                                                                                                                                & Segfault                      \\\hline
    crypto.privateKey.export                         & \verb|kty|                                                                                                                                                & Segfault                      \\\hline
    \multirow{2}{*}{crypto.createPrivateKey}         & \verb|type|                                                                                                                                               & Segfault                      \\\cline{2-3}
                                                     & \verb|passphrase|                                                                                                                                         & Segfault                      \\\hline
    \multirow{2}{*}{crypto.createPublicKey}          & \verb|type|                                                                                                                                               & Segfault                      \\\cline{2-3}
                                                     & \verb|passphrase|                                                                                                                                         & Segfault                      \\\hline
    fetch                                            & {\begin{tabular}[l]{@{}l@{}}\verb|socketPath|, \verb|body|,\\\verb|method|, \verb|referrer|\end{tabular}}                                                 & SSRF                          \\\hline
    fs.createWriteStream                             & \verb|mode|                                                                                                                                               & PE                            \\\hline
    https.get                                        & {\begin{tabular}[l]{@{}l@{}}\verb|hostname|, \verb|headers|,\\\verb|method|, \verb|path|, \verb|port|,\\\verb|NODE_TLS_REJEC...|\end{tabular}}            & SSRF                          \\\hline
    \multirow{2}{*}{https.request}                   & {\begin{tabular}[l]{@{}l@{}}\verb|hostname|, \verb|headers|,\\\verb|method|, \verb|path|, \verb|port|,\\\verb|NODE_TLS_REJEC...|\end{tabular}}            & SSRF                          \\\cline{2-3}
                                                     & \verb|0|                                                                                                                                                  & Segfault                      \\\hline
    http.get                                         & {\begin{tabular}[l]{@{}l@{}}\verb|hostname|, \verb|headers|,\\\verb|method|, \verb|path|, \verb|port|\end{tabular}}                                       & SSRF                          \\\hline
    http.request                                     & {\begin{tabular}[l]{@{}l@{}}\verb|hostname|, \verb|headers|,\\\verb|method|, \verb|path|, \verb|port|\end{tabular}}                                       & SSRF                          \\\hline
    http.Server.listen                               & \verb|backlog|                                                                                                                                            & Segfault                      \\\hline
    import                                           & \verb|source|                                                                                                                                             & ACE                           \\\hline
    require (v18.13.0)                               & \verb|main|                                                                                                                                               & ACE                           \\\hline
    Socket.send                                      & \verb|address|                                                                                                                                            & SSRF                          \\\hline
    stream.Duplex                                    & \verb|readableObjectMode|                                                                                                                                 & Segfault                      \\\hline
    tls.TLSSocket.connect                            & \verb|path|                                                                                                                                               & Segfault                      \\\hline
    vm.SyntheticModule                               & {\begin{tabular}[l]{@{}l@{}}\verb|sourceText|,\\\verb|lineOffset|,\\\verb|columnOffset|\end{tabular}}                                                     & ACE                           \\\hline
    zlib.createGzip().write                          & \verb|writableObjectMode|                                                                                                                                 & Segfault                      \\\hline
  \end{tabular}
  \caption{A summary of the exploitable first-order gadgets in \textbf{Node.js}. \emph{Gadget} identifies the public API that triggers a gadget; \emph{Properties} specifies which properties must be polluted; \emph{Attack Type} specifies one of Arbitrary Code/Command Execution (ACE), Cryptographic Downgrade (CD), Path Traversal (PT), Privilege Escalation (PE), Server Side Request Forgery (SSRF), or Segfault.}
  \label{tab:nodejs-gadgets}
\end{table}

\begin{table}
  \scriptsize
  \centering
  \begin{tabular}{|r|l|l|} 
    \hline
    \textbf{Gadget}                             & \textbf{Properties}                                                                                 & \textbf{Attack Type}        \\\hline

    fetch                                       & \verb|body|, \verb|headers|, \verb|method|, \verb|0|                                                & SSRF                        \\\hline\
    \multirow{8}{*}{Worker}                     & \verb|env|                                                                                          & PE                          \\\cline{2-3}
                                                & \verb|ffi|                                                                                          & PE                          \\\cline{2-3}
                                                & \verb|hrtime|                                                                                       & PE                          \\\cline{2-3}
                                                & \verb|net|                                                                                          & PE                          \\\cline{2-3}
                                                & \verb|read|                                                                                         & PE                          \\\cline{2-3}
                                                & \verb|run|                                                                                          & PE                          \\\cline{2-3}
                                                & \verb|sys|                                                                                          & PE                          \\\cline{2-3}
                                                & \verb|write|                                                                                        & PE                          \\\hline

    \multirow{2}{*}{Deno.makeTempDir}           & \verb|dir|                                                                                          & PT                          \\\cline{2-3} 
                                                & \verb|prefix|                                                                                       & PT                          \\\hline
    \multirow{2}{*}{Deno.makeTempDirSync}       & \verb|dir|                                                                                          & PT                          \\\cline{2-3}
                                                & \verb|prefix|                                                                                       & PT                          \\\hline
    \multirow{2}{*}{Deno.makeTempFile}          & \verb|dir|                                                                                          & PT                          \\\cline{2-3}
                                                & \verb|prefix|                                                                                       & PT                          \\\hline
    \multirow{2}{*}{Deno.makeTempFileSync}      & \verb|dir|                                                                                          & PT                          \\\cline{2-3}
                                                & \verb|prefix|                                                                                       & PT                          \\\hline
    Deno.mkdir                                  & \verb|mode|                                                                                         & PE                          \\\hline
    Deno.mkdirSync                              & \verb|mode|                                                                                         & PE                          \\\hline
    \multirow{3}{*}{Deno.open}                  & \verb|append|                                                                                       & UM                          \\\cline{2-3}
                                                & \verb|mode|                                                                                         & PE                          \\\cline{2-3}
                                                & \verb|truncate|                                                                                     & UM                          \\\hline
    \multirow{3}{*}{Deno.openSync}              & \verb|append|                                                                                       & UM                          \\\cline{2-3}
                                                & \verb|mode|                                                                                         & PE                          \\\cline{2-3}
                                                & \verb|truncate|                                                                                     & UM                          \\\hline
    \multirow{2}{*}{Deno.writeFile}             & \verb|append|                                                                                       & UM                          \\\cline{2-3}
                                                & \verb|mode|                                                                                         & PE                          \\\hline
    \multirow{2}{*}{Deno.writeFileSync}         & \verb|append|                                                                                       & UM                          \\\cline{2-3}
                                                & \verb|mode|                                                                                         & PE                          \\\hline
    \multirow{2}{*}{Deno.writeTextFile}         & \verb|append|                                                                                       & UM                          \\\cline{2-3}
                                                & \verb|mode|                                                                                         & PE                          \\\hline
    \multirow{2}{*}{Deno.writeTextFileSync}     & \verb|append|                                                                                       & UM                          \\\cline{2-3}
                                                & \verb|mode|                                                                                         & PE                          \\\hline
    \multirow{3}{*}{Deno.run}                   & \verb|cwd|                                                                                          & PT                          \\\cline{2-3}
                                                & \verb|gid|                                                                                          & PE                          \\\cline{2-3}
                                                & \verb|uid|                                                                                          & PE                          \\\hline
    \multirow{3}{*}{Deno.Command}               & \verb|cwd|                                                                                          & PT                          \\\cline{2-3}
                                                & \verb|gid|                                                                                          & PE                          \\\cline{2-3}
                                                & \verb|uid|                                                                                          & PE                          \\\hline

    cp.exec                                     & \verb|shell|, \verb|env|                                                                            & ACE                         \\\hline
    cp.execFileSync                             & \verb|shell|, \verb|env|                                                                            & ACE                         \\\hline
    cp.execSync                                 & \verb|shell|, \verb|env|                                                                            & ACE                         \\\hline
    \multirow{3}{*}{cp.spawn}                   & \verb|shell|, \verb|env|                                                                            & ACE                         \\\cline{2-3}
                                                & \verb|gid|                                                                                          & PE                          \\\cline{2-3}
                                                & \verb|uid|                                                                                          & PE                          \\\hline
    cp.spawnSync                                & \verb|shell|, \verb|env|                                                                            & ACE                         \\\hline
    \multirow{2}{*}{fs.appendFile}              & \verb|length|                                                                                       & Loop                        \\\cline{2-3}
                                                & \verb|offset|                                                                                       & OOM                         \\\hline
    \multirow{2}{*}{fs.writeFile}               & \verb|length|                                                                                       & Loop                        \\\cline{2-3}
                                                & \verb|offset|                                                                                       & OOM                         \\\hline
    http.request                                & \verb|hostname|, \verb|method|, \verb|path|, \verb|port|                                            & SSRF                        \\\hline
    https.request                               & \verb|hostname|, \verb|method|, \verb|path|, \verb|port|                                            & SSRF                        \\\hline
    zlib.createBrotliCompress                   & \verb|params|                                                                                       & Panic                       \\\hline

    \multirow{2}{*}{json.JsonStringifyStream}   & \verb|prefix|                                                                                       & UM                          \\\cline{2-3}
                                                & \verb|suffix|                                                                                       & UM                          \\\hline
    log.FileHandler                             & \verb|formatter|                                                                                    & LP                          \\\hline
    \multirow{2}{*}{tar.Tar.append}             & \verb|gid|                                                                                          & PE                          \\\cline{2-3}
                                                & \verb|uid|                                                                                          & PE                          \\\hline
    yaml.stringify                              & \verb|indent|                                                                                       & OOM                         \\\hline
  \end{tabular}
  \caption{A summary of the exploitable first-order gadgets in \textbf{Deno}. \emph{Gadget} identifies the public API that triggers a gadget; \emph{Properties} specifies which properties must be polluted; \emph{Attack Type} specifies one of Arbitrary Code/Command Execution (ACE), Log Pollution (LP), Loop, Out of Memory (OOM), Panic, Path Traversal (PT), Privilege Escalation (PE), Server Side Request Forgery (SSRF), or Unauthorized Modifications (UM).}
  \label{tab:deno-gadgets}
\end{table}

\begin{lstlisting}[caption={Injected snippet for polluting with a string value.},label={lst:simulating-pollution}]
let __pollutedValue = '0xEFFACED', __accessIndex = 0;
Object.defineProperty(Object.prototype, '${prop}', {
 get: function() {
  const returnValue = __pollutedValue + __accessIndex;
  __accessIndex += 1;
  try {
   throw new Error();
  } catch(error) {
   globalThis.log(returnValue + ' source stack: ' + error.stack);
  }
  return returnValue;
 },
 set: function(newValue) {
  Object.defineProperty(this, '${prop}', {
   value: newValue,
   writable: true,
   enumerable: true,
   configurable: true
  });
 },
 enumerable: ${prop === FORIN_SYMBOL ? "true" : "false"},
 configurable: true,
});
\end{lstlisting}

\end{document}